\documentclass[aps,preprint,showpacs,floatfix,nofootinbib,preprintnumbers,superscriptaddress,amsmath,amssymb]{revtex4-1}

\usepackage{bm}
\usepackage{color}
\usepackage{float}
\usepackage{dcolumn}
\usepackage{multirow}
\usepackage[utf8]{inputenc}
\usepackage[english]{babel}
\usepackage[letterpaper, portrait, margin=1in]{geometry}
\usepackage[font={small,it}]{caption}
\usepackage{graphicx}
\usepackage[font={tiny,it}]{subcaption}
\usepackage{nicefrac}
\usepackage{wrapfig}
\usepackage{amsmath}

\usepackage{listings}
\usepackage{color}
\definecolor{dkgreen}{rgb}{0,0.6,0}
\definecolor{gray}{rgb}{0.5,0.5,0.5}
\definecolor{mauve}{rgb}{0.58,0,0.82}
\begin{document}

\title{Accelerating the Convergence of Coupled Cluster Calculations of the Homogeneous Electron Gas Using Bayesian Ridge Regression}

\author{Julie Butler}
\affiliation{Department of Biochemistry, Chemistry, and Physics, University of Mount Union, Alliance, Ohio 46601, USA}
\affiliation{Department of Physics and Astronomy and Facility for Rare Isotope Beams and National Superconducting  Cyclotron Laboratory, Michigan State University, East Lansing, MI 48824, USA} 
\email[Corresponding Author: ]{butlerju@mountunion.edu}
\author{Morten Hjorth-Jensen} 
\affiliation{Department of Physics and Astronomy and Facility for Rare Isotope Beams and National Superconducting  Cyclotron Laboratory, Michigan State University, East Lansing, MI 48824, USA} 
\affiliation{Department of Physics and Center for Computing in Science Education, University of Oslo, N-0316  Oslo, Norway} 
\author{Justin G. Lietz}
\affiliation{National Center for Computational Sciences, Oak Ridge National Laboratory, Oak Ridge, TN 37831, USA}


\maketitle

\section*{Abstract}
The homogeneous electron gas is a system that has many applications in chemistry and physics. However, its infinite nature makes studies at the many-body level complicated due to long computational run times. Because it is size extensive, coupled cluster theory is capable of studying the homogeneous electron gas, but it still poses a large computational challenge as the time needed for precise calculations increases in a polynomial manner with the number of particles and single-particle states. Consequently, achieving convergence in energy calculations becomes challenging, if not prohibited, due to long computational run times and high computational resource requirements. This paper develops the sequential regression extrapolation (SRE) to predict the coupled cluster energies of the homogeneous electron gas in the complete basis limit using Bayesian ridge regression and many-body perturbation theory correlation energies to the second order to make predictions from calculations at truncated basis sizes.  Using the SRE method we were able to predict the coupled cluster doubles energies for the electron gas across a variety of values of N and $r_s$, for a total of 70 predictions, with an average error of 5.20x10$^{-4}$ Hartrees while saving 88.9 hours of computational time. The SRE method can accurately extrapolate electron gas energies to the complete basis limit, saving both computational time and resources. Additionally, the SRE is a general method that can be applied to a variety of systems, many-body methods, and extrapolations.
\newpage

\section*{Introduction}
 A proper understanding of the role of correlations beyond a mean-field
description is of great importance for quantum many-particle
theories. For all possible quantum-mechanical systems, either finite
ones or infinite systems like the homogeneous electron gas in two or
three dimensions or nuclear or neutron star matter, the concept of an
independent particle motion continues to play a fundamental role in
studies of many-particle systems.  Deviations from such an independent
particle, or mean-field, picture have normally been interpreted as a
possible measure of correlations.  The latter are expected to reveal
important features of both the structure and the dynamics of a
many-particle system beyond a mean-field approximation.

The homogeneous electron gas (HEG)
is an infinite matter system composed of electrons with a uniform positive background charge \cite{Ref1, Ref2, Ref3, Ref68, Ref71, Ref72, Ref95}. The ions are stationary and the system as a whole
is neutral. It is widely studied in chemistry and physics because it is a reasonable model for many systems of interest. For instance, the three-dimensional HEG is fundamental in density functional theory's local-density approximation and can represent electrons in alkali metals and doped semiconductors. Two-dimensional electron fluids are 
observed on metal and liquid-helium surfaces, as well as 
at metal-oxide-semiconductor interfaces. 
The HEG  serves also as a foundation for developing many-body frameworks applicable to more complex systems such as infinite nuclear matter.

The HEG in two and three dimensions represent also some of the few examples of systems of many
interacting particles that allow for analytical solutions of the mean-field
Hartree-Fock equations. 
Irrespective of this simplicity, the HEG in both two and
three-dimensions, has eluded a proper description of correlations in
terms of various first principle methods, except perhaps for quantum
Monte Carlo methods. In particular, the diffusion Monte Carlo
calculations of Ceperley {\em et al.}
\cite{ceperley1980,tanatar1989}, are presently still considered the
best possible benchmarks for two- and three-dimensional electron
gas. In Ref.~\cite{ceperley1980},
the fermion sign problem is handled using 
a released-node approximation. Similar accuracy has been 
obtained in more recent calculations using the 
backflow-correlation technique \cite{kwon1998}, which is used 
to relax the simpler fixed-node approximation.  Recently, variational Monte Carlo approaches using deep neural networks \cite{kim2023,cassella2023}
result in ground state energies for a variety of densities which show minima below the best diffusion Monte Carlo results. 
In addition to various Monte Carlo approaches, the electron gas has been studied using a large number of different approaches and many-body methods. Amongst such many-body methods, we have coupled cluster (CC) theory.
Coupled cluster theory is a powerful method used in nuclear physics, condensed matter physics, quantum chemistry, and other fields to predict ground state energy accurately \cite{Ref21, Ref47, Ref70, Ref106, Ref107,Ref147,Ref149,Ref150,Ref151}. It organizes and includes interactions beyond the mean-field in a systematic manner, making it a popular choice for many-body calculations. Although CC truncates interactions, making it less accurate than full configuration interaction theory, it offers faster computational run times. Compared to other many-body methods like many-body perturbation theory, CC achieves significantly higher accuracy, making it an attractive compromise between speed and accuracy.

Singal and Das \cite{singal1973} were the first to study 
the electron gas using a CC approach. The approximation
they used is similar to the Brueckner Hartree-Fock method \cite{haftel1970}, 
and did not properly include particle-hole ring diagrams.
Later, Freeman performed CC calculations \cite{freeman1977} in 
which only ring diagrams and their exchange parts were 
retained in a CC doubles approximation. The results of both
Singal and Das \cite{singal1973} and Freeman 
\cite{freeman1977} compared well with dielectric-function
approaches. These calculations of  
the three-dimensional electron gas were improved by Bishop and
L{\"u}hrmann \cite{Ref72,bishop1982}. Bishop and
L{\"u}hrmann derived a CC SUB2 approximation (also called
CC doubles (CCD) \cite{Ref21}) for the electron gas, 
extended with some ladder contributions from higher-order
amplitudes. As the authors described in 
Ref.~\cite{Ref72}, the CCD approximation 
contains many more diagrammatic classes than
partial-summation techniques derived from perturbation 
theory. The CCD approximation takes account of 
particle-particle and hole-hole ladders, particle-hole
ring diagrams including exchange terms, and many other
diagrams to infinite order in perturbation theory. 
Recently, Shepherd \emph{et al.}
\cite{Ref1,shepherd2013a,shepherd2013b,Ref71} have
taken up again the CC effort for the three-dimensional electron
gas. The CC calculations for the electron gas from the 1970s and 1980s
were all done at the thermodynamic limit
\cite{singal1973,freeman1977,Ref72,bishop1982}. Unfortunately,
singularities related to the Coulomb two-body interaction may
complicate numerical calculations at the thermodynamic limit.
Shepherd \emph{et al.} approximate the electron gas using finite cubic
boxes
\cite{Ref1,shepherd2013a,shepherd2013b,Ref71}. 
However, to make CC calculations feasible in the HEG using finite lattices, truncations are necessary, thereby introducing errors that diminish accuracy. The truncation in the basis of single-particle states ($M$) leads to basis incompleteness errors while truncating the number of electrons ($N$) and the HEG volume ($\Omega$) introduces finite-size errors. One way to reduce these errors is by performing CC calculations at high values of $M$ and $N$, such that the calculations have converged. However, this approach demands extensive computational resources and time. Since basis incompleteness errors and finite size errors arise from different truncations, they can be addressed independently \cite{Ref1, Ref2}. This work focuses on removing the basis truncation errors from CC doubles calculations of the HEG using machine learning.

Machine learning (ML), a field at the intersection of data science and artificial intelligence, has become increasingly prominent as a tool in physics and chemistry, which can be applied to a wide variety of problems \cite{Ref11, Ref12, Ref19, Ref23, Ref6, Ref27, Ref28, Ref32, Ref109, Ref110, Ref112, Ref114, Ref117}. This paper aims to develop an ML-based extrapolation method, called sequential regression extrapolation (SRE) that uses Bayesian ridge regression. This method extrapolates CC correlation energies calculated at smaller M to eliminate basis incompleteness errors. Despite the challenges of applying ML for extrapolation and limited data sets due to computational constraints, the proposed approach is expected to significantly reduce the run time of fully converged CC calculations without sacrificing accuracy.

Previous studies have explored ML techniques in CC calculations, including predicting CC correlation energies from many-body perturbation theory amplitudes and using neural networks to eliminate finite-size errors in nuclear CC calculations \cite{Ref6, Ref7, Ref81}. Traditional extrapolation methods have also been employed to extrapolate converged CC correlation energies for the HEG and other many-body systems, to eliminate basis incompleteness and finite-sized errors \cite{Ref1, Ref2, Ref74, Ref86, Ref94, Ref96, Ref98, Ref99}. However, this paper distinguishes itself by applying an ML-based extrapolation technique, SRE, to predict CC correlation energies using Bayesian ridge regression, specifically targeting the HEG.

The paper's structure begins with a theoretical explanation of the homogeneous electron gas, many-body perturbation theory, and coupled cluster theory. It then introduces the development of the SRE algorithm and its application to accelerate the convergence of CC correlation energies concerning the number of single-particle states and electrons. Finally, the results and discussion section presents the time savings achieved with the SRE algorithm, and the accuracy of extrapolation in terms of single-particle states.
\section*{Background}
Coupled cluster theory (CC) is a highly effective many-body method that approximates the solution to the Schr\"{o}dinger equation to a high degree of accuracy \cite{Ref149, Ref150, Ref151, Ref152, Ref153}. Coupled cluster theory is widely used in quantum chemistry to accurately predict the ground state energy of complex systems \cite{Ref147}. By systematically incorporating interactions beyond the mean field, CC offers a robust framework for many-body calculations.  CC outperforms other many-body methods like many-body perturbation theory in terms of accuracy, making it an appealing choice that strikes a balance between speed and precision. However, even though CC is size extensive (an important quality for its application to infinite matter systems) the polynomial time scaling, with respect to the size of the system, can make CC run times prohibitively long for large systems \cite{Ref157}.

Coupled cluster theory starts with the exponential ansatz that states a many-body wavefunction can be created by applying an exponential cluster operator to the Fermi vacuum state, shown in Eq.~\eqref{eq:expansatz}. The cluster operator consists of N excitation operators, where N is the number of electrons or particles in the system. However, for practical calculations, the number of terms in the cluster operator is often truncated to a set level \cite{Ref140, Ref141, Ref142, Ref143, Ref155, Ref157}. In this work, the focus is on the doubles level (CCD), which approximates the cluster operator with just the two-particle two-hole excitation operator (Eq.~\eqref{eq:amplitudes}). Since the homogeneous electron gas is an infinite matter system, the terms which result from coupled cluster singles are zero due to momentum conservation. Thus performing CCD calculations here are equivalent to performing coupled cluster singles and doubles (CCSD) calculations.

\begin{equation}
\label{eq:expansatz}
    |\Psi\rangle = e^{\hat{T}}|\Phi_0\rangle,
\end{equation}
and
\begin{equation}
    \hat{T} = \sum_{i=1}^N \hat{T}_i \longrightarrow \hat{T}_{CCD} = \hat{T}_2.
    \label{eq:amplitudes}
\end{equation}

A CCD calculation has an expected run time of $O(M^6)$, where $M$ is the number of single-particle states in the calculation \cite{Ref157}. Furthermore, it is important to note that a CCD calculation contains all correlations needed to compute the many-body perturbation theory to the second-order (MBPT2) results, so MBPT2 energies can be computed from a CCD calculation with little additional computational costs. The MBPT2 energy is less accurate than the CCD energy but it has a significantly smaller run time while still incorporating some aspects of the interactions beyond the mean field. 

The energy calculated through CC, denoted as E$_{CC}$, is typically presented as the correlation energy, which is the CC energy minus the reference energy obtained from Hartree-Fock theory
\begin{equation}
    E_{CC} = \langle \Phi_0 | e^{-\hat{T}} \hat{H} e^{\hat{T}} | \Phi_0 \rangle = E_{REF} + \Delta E_{CC}.
\end{equation}

MBPT2 energies can be found, as shown in Eq.~\eqref{mbpt}, are are often presented as correlation energies as well.

\begin{equation}\label{mbpt}
    E_{MBPT} = E_{REF} + \frac{1}{4} \sum_{ijab} \frac{\langle ij | \hat{v}_2 | ab \rangle \langle ab | \hat{v}_2 | ij \rangle}{(\epsilon_i + \epsilon_j) - (\epsilon_a + \epsilon_b)}
\end{equation}

In the above equation, $E_{REF}$ is the reference energy obtained from Hartree-Fock theory, $\epsilon_p$ is the energy of state p, and $\langle ij | \hat{v}_2 | ab \rangle$ is the two-body operator between the filled states i and j, and the empty states a and b.

As mentioned above, the HEG is an infinite matter system consisting of electrons with a uniform positive background charge, and studies are important in diverse areas of chemistry and physics \cite{Ref1, Ref2, Ref69, Ref71, Ref72}. In the context of the three-dimensional HEG, it plays a fundamental role in the local-density approximation of density functional theory. Additionally, the HEG can effectively model electrons in alkali metals and doped semiconductors and provides a foundation for developing many-body frameworks applicable to more intricate infinite matter systems, including infinite nuclear matter \cite{Ref1, Ref2, Ref3, Ref71, Ref72}.

The HEG Hamiltonian for $N$ electrons includes kinetic energy, electron-electron interactions, and the Madelung term. The last two terms make up the Ewald interaction, which separates short-range and long-range components of the Coulomb interaction \cite{Ref1, Ref2}. The Hamiltonian is defined as:

\begin{equation}
\hat{H} = -\frac{1}{2}\sum_\alpha \nabla^2_\alpha + \frac{1}{2}\sum_{\alpha \neq \beta}\hat{v}_{\alpha\beta} + \frac{1}{2}Nv_M,
\end{equation}
where $\alpha$ and $\beta$ are electron indices. The two-electron operator, $\hat{v}_{\alpha\beta}$, accounts for the contributions from the electron-electron interactions. 

The one-electron basis set consists of plane waves, which are given by the wave function:

\begin{equation}
\psi_j(\vec{r},\sigma) = \sqrt{\frac{1}{\Omega}}\ e^{i\vec{k}j\cdot\vec{r}}\ \delta_{\sigma_j,\sigma},
\end{equation}
where $\Omega$ is the volume of the HEG, $\vec{k}_j$ is the wave vector, and $\delta{\sigma_j,\sigma}$ is the Kronecker delta \cite{Ref3,Ref13}. The density of the HEG is often described by the Wigner-Seitz radius, $r_s$, which is related to the size of the sphere that contains the electron gas. In this work, we are interested in high-density HEG systems and will restrict $r_s < 1.0$ \cite{Ref91, Ref92}. 

When applying CC to the HEG, certain truncations become necessary to make calculations computationally feasible, albeit at the cost of introducing errors that compromise accuracy \cite{Ref1, Ref2, Ref71, Ref72}. The truncation of single-particle states ($M$) gives rise to basis incompleteness errors while truncating electrons ($N$) and the corresponding volume of the HEG ($\Omega$) introduces finite-size errors while also necessitating the inclusion of boundary conditions (periodic boundary conditions are used in this work). Finally, the CC cluster operator must also be truncated, as discussed previously. 

This paper focuses on the truncation of $M$, which introduces basis truncation errors into the calculations compared to calculates performed at the complete basis limit ($M \longrightarrow \infty$). One approach to mitigate this error involves performing CC calculations at high values of $M$. However, this strategy imposes significant computational demands and time requirements, since CCD computational costs scale as $O(M^6)$ \cite{Ref157}. For the calculations performed in this work, the energies converted with respect to basis size at M=6,142 single-particle states, making the computational costs of $O(M^6)$ a very unattractive prospect. This paper aims to address the truncation errors in coupled cluster calculations by proposing a method that retains accuracy without relying on computationally burdensome calculations at larger $M$ values. The focus is on removing the basis incompleteness error by performing calculations at smaller values of $M$ and extrapolating the coupled cluster (CC) energy results to achieve convergence in the complete basis limit.

Specifically, the method involves conducting calculations at a fixed number of electrons, N, and a fixed Wigner-Seitz radius (r$_s$) using 15 increasing, yet relatively small, values of $M$ ($M \leq$ 2,090). Note that the exact values of $M$ used in the calculations depend on the number of particles in the system. A machine learning-based algorithm is employed to extrapolate the CC energy to convergence. The resulting converged CC energy should approximate the CC energy obtained at a high $M$ value ($M = 6,142$ in this work), effectively mitigating the basis incompleteness error, while saving significant computational time and resources.

\section*{Methodology}
\begin{figure}
     \begin{subfigure}[b]{0.32\textwidth}
         \centering
         \includegraphics[width=\textwidth]{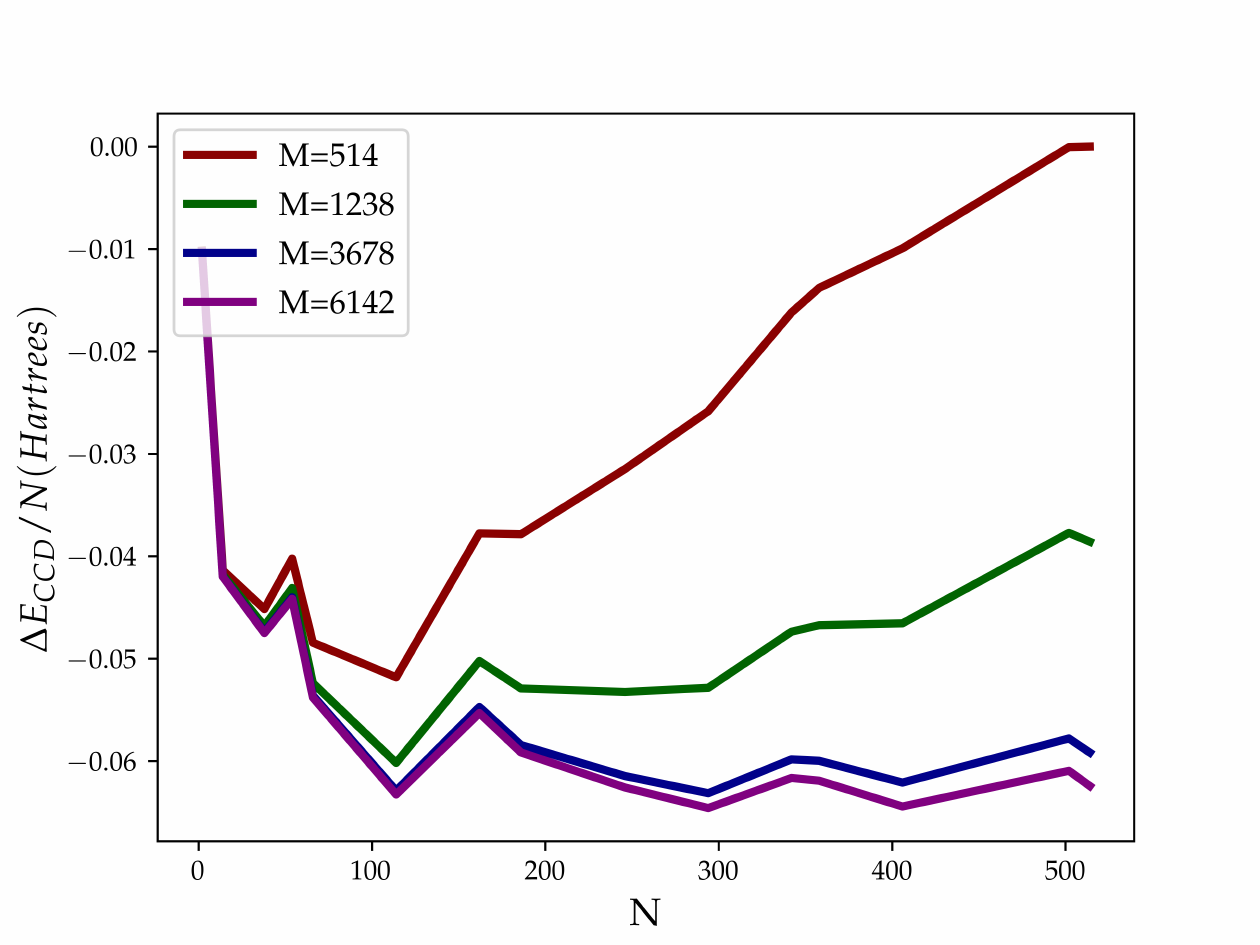}
        \caption{$\Delta E_{CCD}$ per electron plotted against $N$ with calculations performed at four different values of $M$. As $M$ increases, the error in the calculation decreases, but the run time drastically increases.}
         \label{fig:increase m}
     \end{subfigure}
     \hfill
          \begin{subfigure}[b]{0.32\textwidth}
         \centering
         \includegraphics[width=\textwidth]{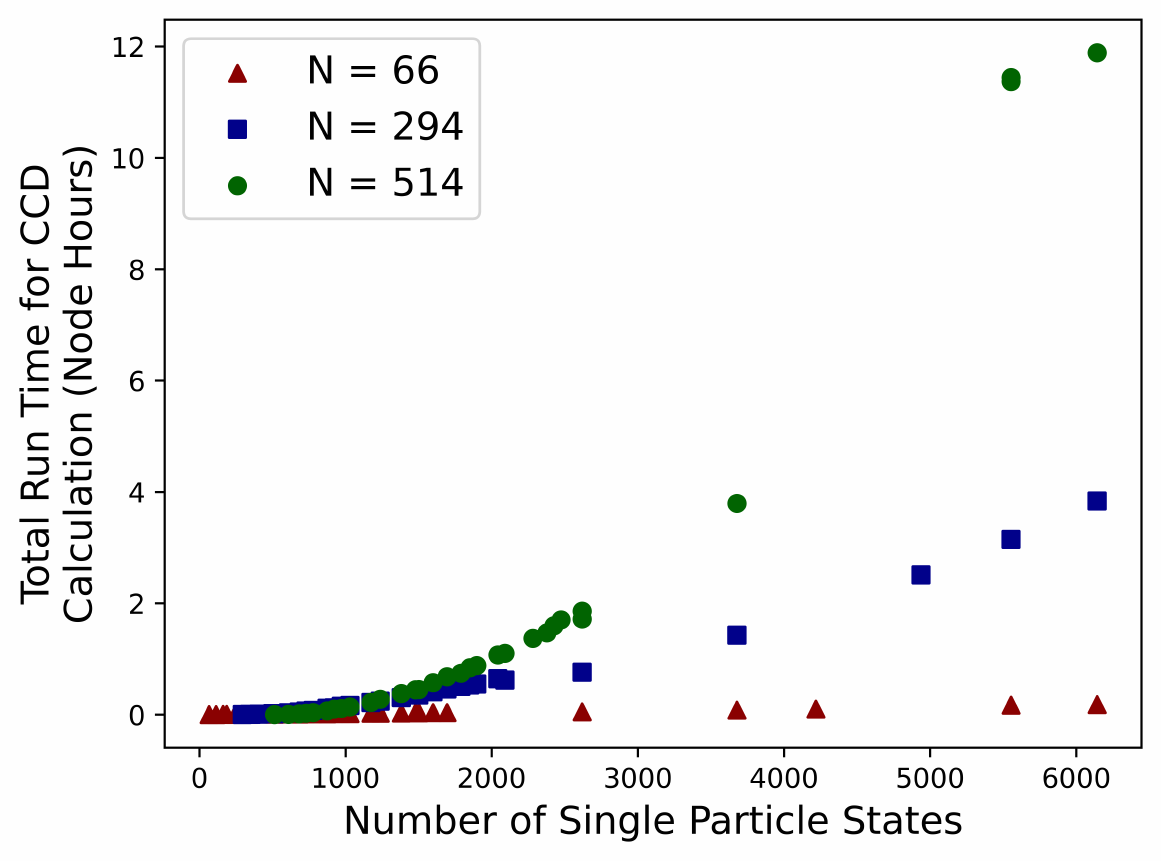}
         \caption{The run times (measured in node hours) for CCD calculations of the HEG as a function of M and at three different values of N.}
         \label{fig:timing}
     \end{subfigure}
     \hfill
          \begin{subfigure}[b]{0.32\textwidth}
         \centering
         \includegraphics[width=\textwidth]{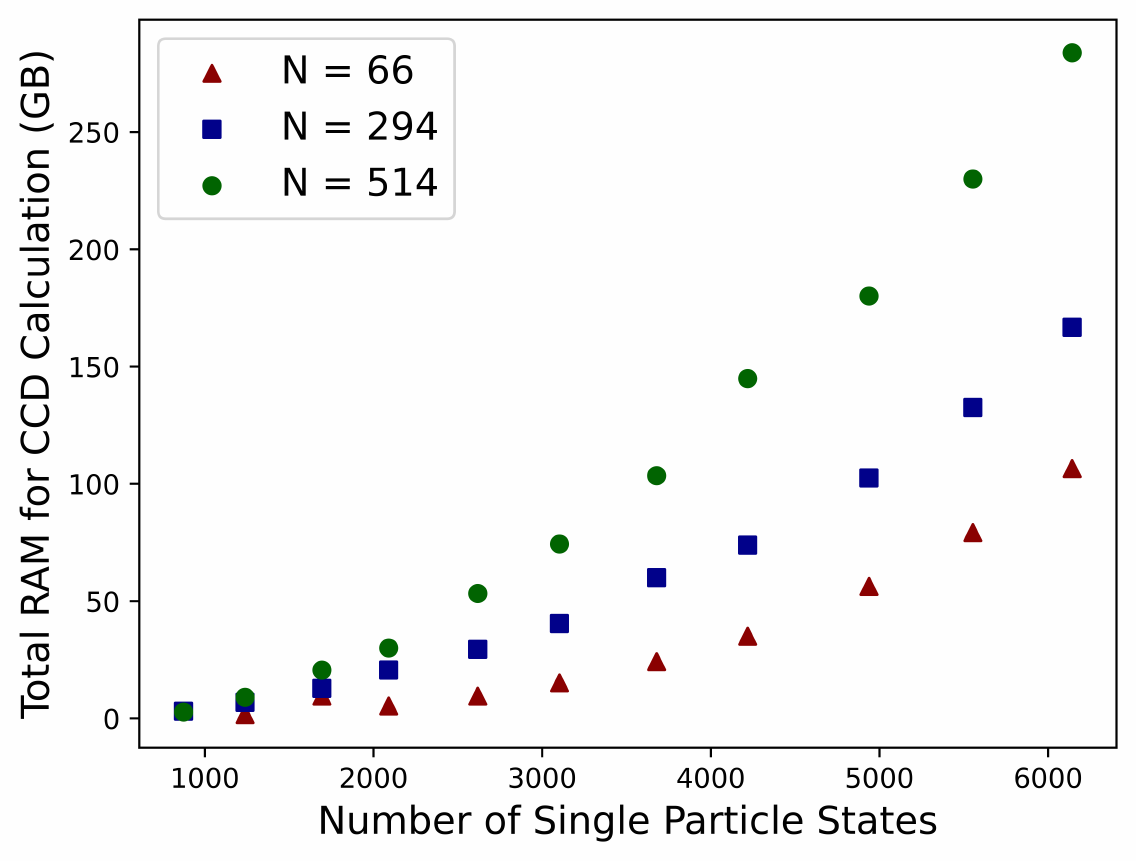}
         \caption{The RAM required for CCD calculations of the HEG as a function of M and at three different values of N.}
         \label{fig:ram}
     \end{subfigure}
     \hfill
     \newline
     \begin{subfigure}[b]{0.32\textwidth}
         \centering
         \includegraphics[width=\textwidth]{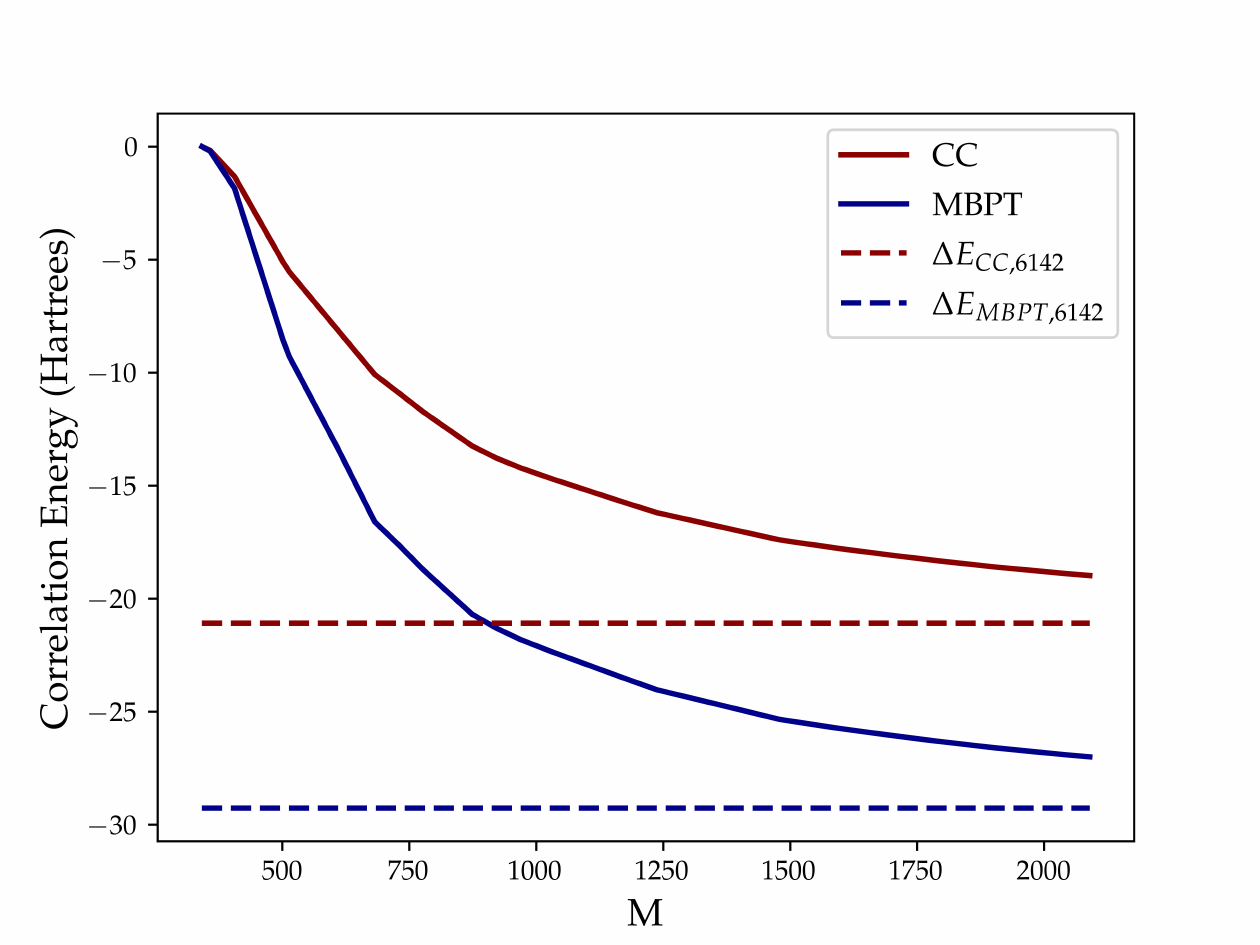}
         \caption{A plot of $\Delta E_{CCD}$ and $\Delta E_{MBPT2}$ versus $M$ for a HEG system with $N=342$ and r$_s$=0.5.  The converged correlation energies calculated at $M=6,142$ are shown with dashed lines for each method.}
         \label{fig:mbpt and cc}
     \end{subfigure}
     \hfill
     \begin{subfigure}[b]{0.32\textwidth}
         \centering
         \includegraphics[width=\textwidth]{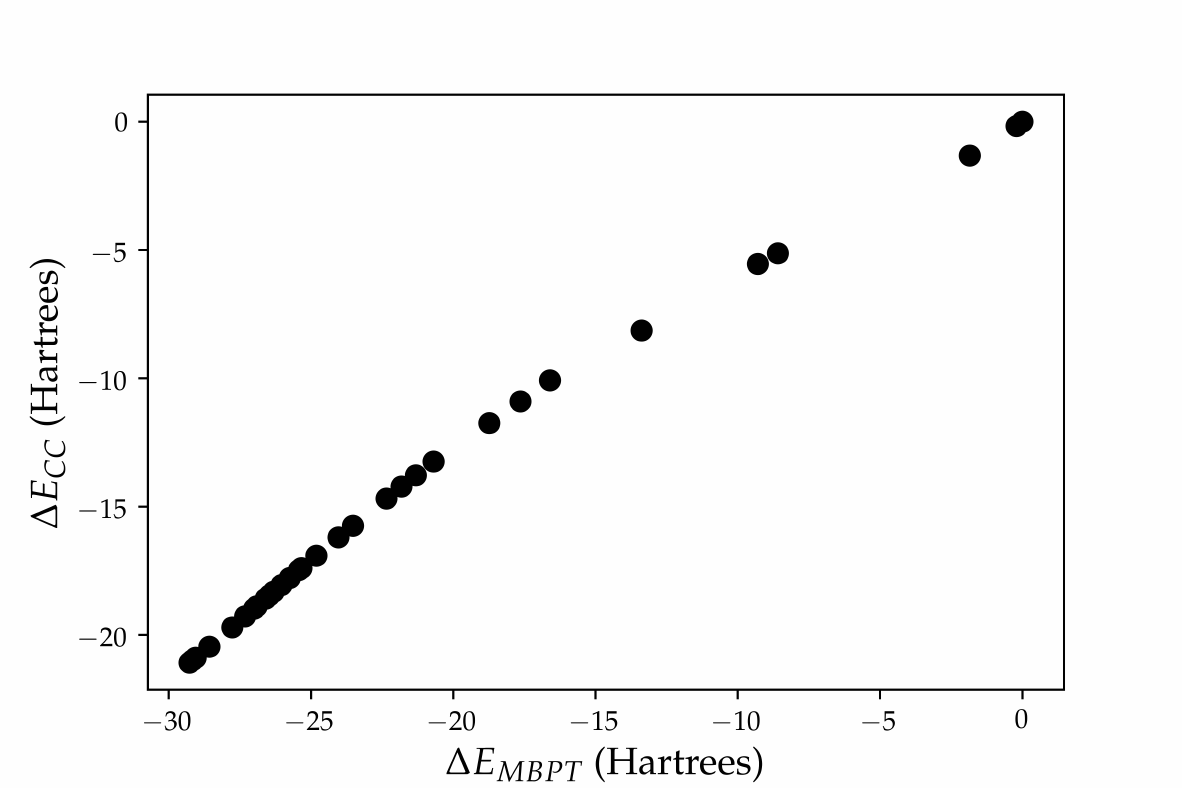}
         \caption{Figure \ref{fig:mbpt and cc} is transformed such that $\Delta E_{CCD}$ are plotted as a function of $\Delta E_{MBPT2}$ (such that each point has the same $M$).  The resulting plot has a linear relationship that becomes stronger to the left (as $M$ increases).}
         \label{fig:mbpt vs. cc}
     \end{subfigure}
     \hfill
     \begin{subfigure}[b]{0.32\textwidth}
         \centering
         \includegraphics[width=\textwidth]{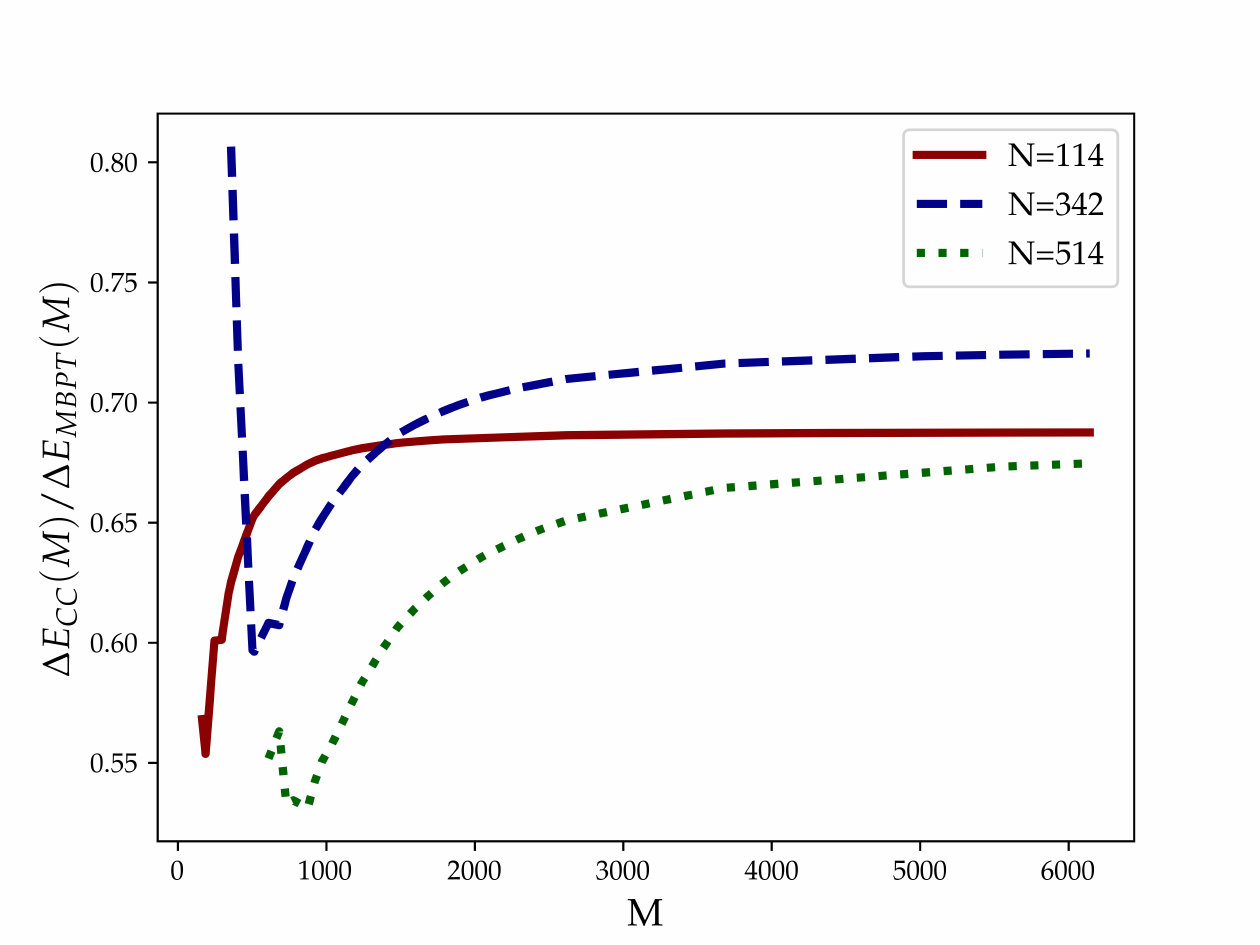}
         \caption{A plot of the ratio of $\Delta E_{CCD}$ to $\Delta E_{MBPT2}$ as a function of $M$ for $N=114$, $342$, and $502$ and r$_s$=0.5.  As $M$ increases, the ratio (or the slope of the line in Figure \ref{fig:ram} becomes increasingly constant. However, it is important to note that the run time increases drastically as $M$ increases.}
         \label{fig:slopes}
     \end{subfigure}
     \hfill 
        \caption{An analysis of the relationship between $\Delta E_{CCD}$ and $\Delta E_{MBPT2}$ for calculations on the HEG at various $M$, $N$, and r$_s$.}
        \label{fig: correlation analysis}
\end{figure}

In this work, the CCD and MBPT2 correlation energies will be denoted as $\Delta$E$_{CCD, M}$ and $\Delta$E$_{MBPT2, M}$, respectively, to show the number of single-particle states in the calculations. Figure \ref{fig:increase m} shows a plot of $\Delta E_{CCD}$ per particle for all magic numbers between two and 502(inclusive), with the CC calculations performed at four different values of $M$ and at $r_s$ = 0.5. The calculations performed at $M=514$ (the red line in the plot) are the least accurate since the data was generated with the lowest value of $M$. When $M=6,142$, the correlation energies are converged (to $10^{-7}$ Hartrees), making this data set (plotted in purple) the most accurate. However, as $M$ increased, the run time and computational resources of the calculation drastically increased as well. 

The computational run times for CCD calculations for the HEG are shown in Figure \ref{fig:timing}, plotted as a function of the number of single-particle states in the calculation for three values of $N$ and $r_s$ = 0.5. Note that the run times are reported in node hours, calculated by multiplying the number of nodes used in the program by the total run time (in hours). This is done to more accurately reflect the computational resources needed by a parallelized program. All calculations were performed on Michigan State University's High Performance Computing Center (HPCC) using Intel Xeon processors with a clock speed of 2.4 GHz and 240 GB of RAM.  All calculations were parallelized using four MPI nodes and 28 OpenMP threads. Run time increases drastically as both $N$ and $M$ increase, with the calculations at higher values of $M$ and $N$ having run times of over twelve node hours. Studies of the homogeneous electron gas frequently require calculations at many values of $N$ and $r_s$, so having one calculation take hours can make large-scale studies prohibitive. 

Figure \ref{fig:ram} shows the RAM needed for CCD calculations for the HEG as a function of $M$ and for three values of $N$ at $r_s$ = 0.5.  Note that these RAM measurements have been taken from a naive CCD code which has no memory optimizations, to show the worst-case scenario in these calculations.  This increase in RAM as $M$ increases is due to an increase in the number of matrix elements as both $N$ and $M$ increase.  Naively, there are ${M \choose N}$ matrix elements, but computational advancements, such as the block-diagonal memory structure described in \cite{Ref5}, do reduce the total number of matrix elements that need to be calculated and stored at one time and since CCD is already truncated, its memory requirements are only $O(M^4)$, which is still quite high for when performing calculations in the complete basis limit

Therefore, this paper aims to predict the converged $\Delta E_{CCD}$ for the HEG using only data calculated at smaller values of $M$, where the computational time and resource requirements are less significant. This paper will use $\Delta E_{CCD,6142}$ as the actual, or most accurate result, and $\Delta E_{CDC}$ has been confirmed to have converged at this point, meaning that the correlation energies have converged to within $10^{-7}$ Hartrees when adding additional single-particle states to the calculations.

Figure \ref{fig:mbpt and cc} shows the convergence of $\Delta E_{CCD}$ as a function of $M$ for a HEG with $N=342$ and r$_s$=0.5. Though the apparent route would be to attempt to predict the converged correlation energy (plotted with a red dashed line) using the converging correlation energies from smaller $M$ calculations, in practice, predictions using this method do not have an acceptable accuracy. However, if we plot $\Delta E_{MBPT2}$ for the same system in addition to $\Delta E_{CCD}$, we see that they follow a very similar pattern. Since $\Delta E_{MBPT2}$ can be generated quickly, the goal of this work is to create a model through which the converged $\Delta E_{CCD}$ can be predicted from $\Delta E_{MBPT2}$. Figure \ref{fig:mbpt vs. cc} is a transformation of Figure \ref{fig:mbpt and cc}, where $\Delta E_{CCD}$ are plotted as a function of $\Delta E_{MBPT2}$. Each point represents a calculation performed with the same value of $M$. When the data is formatted in this way, $M$ increases from right to left. As the graph moves from right to left, it becomes increasingly linear (i.e., the ratio $\Delta E_{CCD}/\Delta E_{MBPT2}$ becomes constant as $M$ increases).

We can see this trend more clearly if we plot $\Delta E_{CCD}/\Delta E_{MBPT2}$ for various $M$ values and $N = 114$, $342$, and $502$ with r$_s$ = 0.5 in Figure \ref{fig:slopes}. For each of these data sets at larger values of $M$, the ratio leveled off to a constant, or near constant, value. Therefore the goal of the research is to be able to predict the constant value of $\Delta E_{CCD}/\Delta E_{MBPT2}$, which occurs at larger values of $M$ by extrapolating from data points collected at lower values of $M$.  We will use this pattern to create the input to a method created by the authors known as sequential regression extrapolation (SRE), a regression-based machine learning algorithm used to accurately extrapolate small data sets generated from many-body codes.

Both $\Delta E_{CCD}$ and $\Delta E_{MBPT2}$ converge as M increases. Therefore, there must be a large value of M where their ratio becomes a constant:

\begin{equation}
\frac{\Delta E_{CCD,Large\ M}}{\Delta E_{MBPT2,Large\ M}} \longrightarrow constant.
\end{equation}

Ridge regression, a supervised machine learning algorithm, will be used to find this constant value using only data collected at low values of M \cite{Ref11, Ref12}. where the ratio is not yet constant. $\Delta E_{CCD}$ and $\Delta E_{MBPT2}$ were calculated for a range of N and r$_s$ values, and for 15 values of M for each data set (2 $\leq$ M$\leq$ 2090, exact values depend on N), resulting in 15 points per data set. For each data set, the training data for the regression algorithm was created by dividing $\Delta E_{CCD}$ by $\Delta E_{MBPT2}$ at the same value of M.

\begin{equation}
y =\frac{\Delta E_{CCD,M}}{\Delta E_{MBPT2,M}} = \frac{\Delta E_{CCD,M_1}}{\Delta E_{MBPT2,M_1}}, \frac{\Delta E_{CCD,M_2}}{\Delta E_{MBPT2,M_2}}, \frac{\Delta E_{CCD,M_3}}{\Delta E_{MBPT2,M_3}}, ..., \frac{\Delta E_{CCD,M_{15}}}{\Delta E_{MBPT2,M_{15}}}
\end{equation}

Next, a ridge regression algorithm with Bayesian hyperparameter tuning is trained on this data set \cite{Ref11, Ref161, Ref162}. The form of the Bayesian ridge regression, $f_{Ridge}$ in Eq. \ref{f ridge}, are defined in Refs. \cite{Ref161} and \cite{Ref162}. However, the way this algorithm is trained is unique because there is no x component of the data set. Instead, the input to the algorithm is two sequential ratios, taking at neighboring values of M, and the output is the ratio at the next value of M. This method of data formatting combined with a regression algorithm gives rise to the SRE algorithm. There are, however, a few important points to note in regards to how the SRE algorithm is trained and used. First, the algorithm is only trained for a given system with set parameters; thus each data point in \ref{fig:fig2} represents a separate training and prediction of the SRE algorithm. The values of $M$ chosen for the training data are taken to correspond to closed energy shells, with the exact values of $M$ ranging between one and 15 open shells with the exact values of $M$ depending on the number of electrons (over the full energy shells which hold the electrons in the ground state). Thus,  the values of $M$ used in the calculations will vary between calculations with different numbers of electrons, though changing the density does of affect the values of $M$ used in the calculation.

\begin{equation}\label{f ridge}
f_{Ridge}(\frac{\Delta E_{CCD,k-2}}{\Delta E_{MBPT2,k-2}}, \frac{\Delta E_{CCD,k-1}}{\Delta E_{MBPT2,k-1}}) = \frac{\Delta E_{CCD,k}}{\Delta E_{MBPT2,k}}
\end{equation}

When Eq. \ref{f ridge} is trained using 15 training points at different values of M, the time-series like formatting results in eleven training points: (($M_1$, $M_2$), $M_3$), (($M_2$, $M_3$), $M_4$), (($M_3$, $M_4$), $M_5$), ... . Once trained, the SRE algorithm is then used to extrapolate this data set to many points until the ratio of correlation energies has converged (for 50 points in total). This value can then be taken as the slope of the graph created with $\Delta E_{CCD}$ plotted as a function of $\Delta E_{MBPT2}$ at high M (Figure \ref{fig:mbpt vs. cc}).

\begin{equation}\label{slope}
\lim_{k\to\infty} \frac{\Delta E_{CCD,k}}{\Delta E_{MBPT2,k}} = slope = m
\end{equation}

Finally, $\Delta E_{MBPT2, 6142}$ is generated, a process that takes less than one second (not including the time to generate the matrix elements). $\Delta E_{MBPT2, 6142}$ is multiplied by the slope, m, to approximate $\Delta E_{CCD, 6142}$.

\begin{equation}
m\Delta E_{MBPT2,6142} \approx \Delta E_{CCD,6142}
\end{equation}

This process is performed for many combinations of N and $r_s$, but each training and prediction takes an average of 30 ms and thus does not contribute significantly to the run time. It is important to note that the data sets used to train a machine learning algorithm in this work consist of 15 points each, making these data sets some of the smallest used in physics applications of machine learning. Many machine learning algorithms need more points to be accurately trained, even up to 1-2 orders of magnitude more, as with some neural networks \cite{Ref6, Ref11, Ref12, Ref17}. It is also of note that machine learning is typically only used to make extrapolations in exceptional cases such as recurrent neural networks. When asked to make predictions outside their training range, many machine learning algorithms perform poorly without significant modifications. However, it will be shown that SRE can make accurate extrapolations using only a small training set, making it a unique machine learning-based algorithm. 

In the analysis of the results and their stability and also range of applicability, one should keep in mind that many-body perturbation theory and coupled cluster theories are theories which explicitly include energy denominators defined in terms of the differences between states below and above the Fermi level. When the system (and this applies to both finite and infinite systems like the electron gas studied here) becomes more dilute, which corresponds to larger values of $r_s$, these energy gaps become smaller and smaller
and result eventually in a worsened convergence and/or unstable results as function of the number of single-particle states. For larger values of $r_s$ one needs more complicated correlations than those included here in order to obtain stable results. For values of $r_s$ larger than those used in this work, many-body perturbation theory to second order results normally in less table results as function of intermediate single-particle states, and limits thus the applicability (with its simplicity defined by second-order MBPT) of the present method.

\section*{Results and Discussion}
The SRE method must meet two metrics to be a useful extrapolator. First, the total time to generate the training data, train the SRE algorithm, and predict the converged correlation energy must be less than the time to do the total calculation for the correlation energy at $M = 6,142$ (significantly less so to be an attractive alternative). Secondly, the accuracy between the predicted converged correlation energies and the calculated converged correlation energies must be high.

 The time needed to train the SRE algorithm and predict the converged correlation energy is on the order of milliseconds, so the primary time savings will come from calculating the training data versus the time needed to perform the calculation at $M = 6,142$. When the number of electrons $N$ is small, the time savings is also small, only 0.024 node hours for $N=66$ and $r_s$ = 0.5. However, the time savings grows quickly as $N$ increases. For example, at $N=246$ and $r_s$ = 0.5, using the SRE method saves 2.05 node hours, and by N=502 and $r_s$ = 0.5, the SRE time savings is 5.88 node hours. Therefore, the SRE method can drastically reduce the run times of predicting the converged correlation energies for the HEG system, especially for studies requiring many values of r$_s$ and $N$.  Additionally, the total time saved from just generating the training data compared to performing all calculations at $M = 6,142$ for all data points shown in Figure \ref{fig:fig2} is 88.9 node hours.

The amount of RAM needed to generate the training points with the highest $M$ value ($M = 2,090$) is only 33.15 GB at 502 electrons (represented the largest combination of $N$ and $M$ across all of the training data sets) while it takes 106.48 GB of RAM just to generate the fully converged coupled cluster correlation energy at 66 electrons (one of the smaller numbers of electrons studied).  At its worst, calculating the fully converged CCD correlation energy at 502 electrons (the highest number of electrons in this study) took 444.48 GB of RAM.  Thus, using the SRE method and only needing to generate a small amount of training data at low values of M saves significant computational time and resources when compared to performing the fully converged calculations at a high value of $M$.

While a reduced run time and lessened computational resources are important, an even more important factor is the accuracy of the predictions.  In Figure \ref{fig:fig2}, the predicted converged $\Delta E_{CCD}$ per electron is plotted with $\Delta E_{CCD,6142}/N$ for a variety of values of $N$ and r$_s$. We limit ourselves to values of $r_s < 1.0$ as we are interested in studying a high-density electron gas \cite{Ref91, Ref92}. The solid lines graph the fully converged calculations performed at $M = 6,142$, the triangular markers plot the SRE prediction for the converged CCD correlation energy per electron at each combination of $N$ and $r_s$, and the shaded regions show the uncertainty of the ML prediction, computed by the Bayesian ridge regression algorithm.  Each color in the plot represents a different value of $r_s$, as shown in the legend of Figure \ref{fig:fig2}. 

\begin{center}
    \includegraphics[scale=0.5]{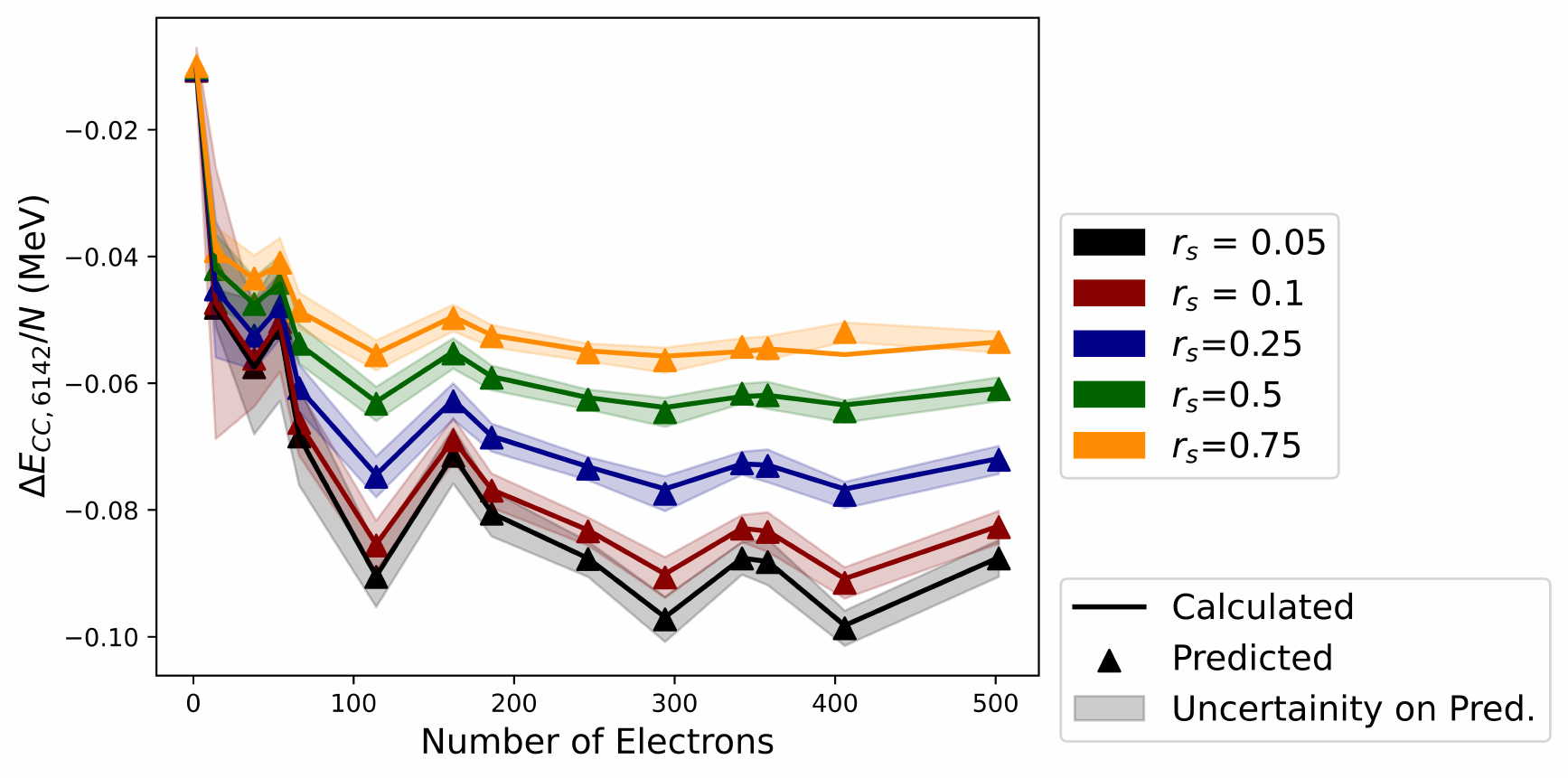}
    \captionof{figure}{The results from performing an SRE extrapolation for various numbers of electrons and values of r$_s$.  Results are plotted against $\Delta E_{CCD,6142}$ and the uncertainties on the predictions are also shown with shaded regions.}
    \label{fig:fig2}
\end{center}

The machine learning predictions lead to an average error for each $r_s$ shown in Table 1, and an overall percent error of 0.39$\%$. If instead of considering the correlation energy, we instead consider transforming the predictions to the full coupled cluster energy ($E_{CCD} = E_0 + \Delta E_{CCD}$), the average percent error drops to 0.011$\%$. Regardless of whether the energies or correlation energies are the desired result, the SRE method is able to make all predictions shown in Figure \ref{fig:fig2} (70 predictions) with an average error of 5.20x10$^{-4}$ Hartrees, meaning this error would not be significant in most coupled cluster calculations. 

\begin{center}

\begin{table}[ht]
    \centering
    \begin{tabular}{|c|c|c|}\hline
         \textbf{r}$_s$ & \textbf{Average Error on Prediction (Hartrees)} & \textbf{Average Uncertainty (Hartrees)}  \\ \hline
         0.05 & 1.59x10${-4}$ & 5.59x10$^{-3}$ \\ \hline
         0.1 & 2.49x10$^{-4}$ & 5.10x10$^{-3}$  \\ \hline
         0.25 & 3.57x10$^{-4}$ &  3.48x10$^{-3}$ \\ \hline
         0.5 & 3.99x10$^{-4}$ &  2.23x10$^{-3}$ \\ \hline
         0.75 & 9.93x10$^{-3}$ & 2.21x10$^{-3}$ \\ \hline

    \end{tabular}
    \caption{A quantitative summary of the SRE prediction accuracy. The first column is the Wigner-Seitz radius, the second the average error (in Hartrees) for each value of $r_s$, and the third column is the average uncertainty on the predictions, taken from the Bayesian ridge regression algorithm.}
    \label{tab:my_label}
\end{table}
\end{center}    

\section*{Conclusion}
Truncating the number of single-particle states in coupled cluster calculations of the homogeneous electron gas is required due to restrictions on computational time and resources.  However, this truncation introduces basis incompleteness errors into the calculations, which reduces their accuracy. The SRE method allows the coupled cluster double correlation energies at the limit of high M to be recovered using CCD correlation energies at small values of $M$.  This saves significant computational time and resources while still achieving an acceptable level of accuracy. To frame the SRE method in another method, it is a dimensionality reduction algorithm for CCD calculations since it allows the number of single-particle states in a calculation to be drastically reduced while still achieving the accuracy of a calculation with a much higher dimension (i.e. the complete basis limit).

Though this work presents the SRE method in the light of performing basis completeness extrapolations for coupled cluster calculations of the homogeneous electron gas, SRE is an extremely general method that can be applied to many problems. With no modifications, this method can perform basis completeness extrapolations for coupled cluster calculations of many systems, both finite and infinite. Of particular interest to the authors is the potential ability of the SRE method to predict the CC correlation energy in the complete basis limit for infinite nuclear matter and atomic nuclei \cite{Ref8, Ref9, Ref4, Ref16, Ref20, Ref43}. Additionally, basis truncations are not unique to coupled cluster calculations, making SRE applicable to other many-body methods as well. Other many-body methods, such as full configuration interaction and in-medium similarity renormalization group theory, may also benefit from an SRE approach \cite{Ref48, Ref65}. Finally, with minor modifications to the set-up of the problem, the SRE method can be used to perform other extrapolations such as extrapolating to the thermodynamic limit to remove finite size effects and extrapolating with respect to the order of a calculation to include higher-order terms without increasing run times, relevant for many-body perturbation theory and coupled cluster theory among others. Thus, the SRE method, capable of accurately performing extrapolations with small data sets, has the possibility of decreasing computational run times and resource requirements of calculations throughout the field of many-body physics, making large-scale studies and studies of complex systems much more feasible.

All code and data needed to recreate the plots and calculations in this paper are available at www.github.com/butler-julie/HEG-SRE.
\section*{Acknowledgements}
This research is supported by the U.S. National Science Foundation Grants No. PHY-2310020 and PHY-2013047. This material is based upon work supported by the U.S. Department of Energy, Office of Science, Office of Workforce Development for Teachers and Scientists, Office of Science Graduate Student Research (SCGSR) program. The SCGSR program is administered by the Oak Ridge Institute for Science and Education (ORISE) for the DOE. ORISE is managed by ORAU under contract number DE-SC0014664. All opinions expressed in this paper are the author’s and do not necessarily reflect the policies and views of DOE, ORAU, or ORISE.
\bibliography{main}

\begin{thebibliography}{76}%
\makeatletter
\providecommand \@ifxundefined [1]{%
 \@ifx{#1\undefined}
}%
\providecommand \@ifnum [1]{%
 \ifnum #1\expandafter \@firstoftwo
 \else \expandafter \@secondoftwo
 \fi
}%
\providecommand \@ifx [1]{%
 \ifx #1\expandafter \@firstoftwo
 \else \expandafter \@secondoftwo
 \fi
}%
\providecommand \natexlab [1]{#1}%
\providecommand \enquote  [1]{``#1''}%
\providecommand \bibnamefont  [1]{#1}%
\providecommand \bibfnamefont [1]{#1}%
\providecommand \citenamefont [1]{#1}%
\providecommand \href@noop [0]{\@secondoftwo}%
\providecommand \href [0]{\begingroup \@sanitize@url \@href}%
\providecommand \@href[1]{\@@startlink{#1}\@@href}%
\providecommand \@@href[1]{\endgroup#1\@@endlink}%
\providecommand \@sanitize@url [0]{\catcode `\\12\catcode `\$12\catcode `\&12\catcode `\#12\catcode `\^12\catcode `\_12\catcode `\%12\relax}%
\providecommand \@@startlink[1]{}%
\providecommand \@@endlink[0]{}%
\providecommand \url  [0]{\begingroup\@sanitize@url \@url }%
\providecommand \@url [1]{\endgroup\@href {#1}{\urlprefix }}%
\providecommand \urlprefix  [0]{URL }%
\providecommand \Eprint [0]{\href }%
\providecommand \doibase [0]{http://dx.doi.org/}%
\providecommand \selectlanguage [0]{\@gobble}%
\providecommand \bibinfo  [0]{\@secondoftwo}%
\providecommand \bibfield  [0]{\@secondoftwo}%
\providecommand \translation [1]{[#1]}%
\providecommand \BibitemOpen [0]{}%
\providecommand \bibitemStop [0]{}%
\providecommand \bibitemNoStop [0]{.\EOS\space}%
\providecommand \EOS [0]{\spacefactor3000\relax}%
\providecommand \BibitemShut  [1]{\csname bibitem#1\endcsname}%
\let\auto@bib@innerbib\@empty
\bibitem [{\citenamefont {Shepherd}\ \emph {et~al.}(2012{\natexlab{a}})\citenamefont {Shepherd}, \citenamefont {Gr\"uneis}, \citenamefont {Booth}, \citenamefont {Kresse},\ and\ \citenamefont {Alavi}}]{Ref1}%
  \BibitemOpen
  \bibfield  {author} {\bibinfo {author} {\bibfnamefont {J.~J.}\ \bibnamefont {Shepherd}}, \bibinfo {author} {\bibfnamefont {A.}~\bibnamefont {Gr\"uneis}}, \bibinfo {author} {\bibfnamefont {G.~H.}\ \bibnamefont {Booth}}, \bibinfo {author} {\bibfnamefont {G.}~\bibnamefont {Kresse}}, \ and\ \bibinfo {author} {\bibfnamefont {A.}~\bibnamefont {Alavi}},\ }\href {\doibase 10.1103/PhysRevB.86.035111} {\bibfield  {journal} {\bibinfo  {journal} {Phys. Rev. B}\ }\textbf {\bibinfo {volume} {86}},\ \bibinfo {pages} {035111} (\bibinfo {year} {2012}{\natexlab{a}})}\BibitemShut {NoStop}%
\bibitem [{\citenamefont {Shepherd}(2016)}]{Ref2}%
  \BibitemOpen
  \bibfield  {author} {\bibinfo {author} {\bibfnamefont {J.~J.}\ \bibnamefont {Shepherd}},\ }\href {\doibase 10.1063/1.4958461} {\bibfield  {journal} {\bibinfo  {journal} {The Journal of Chemical Physics}\ }\textbf {\bibinfo {volume} {145}},\ \bibinfo {pages} {031104} (\bibinfo {year} {2016})},\ \Eprint {http://arxiv.org/abs/https://doi.org/10.1063/1.4958461} {https://doi.org/10.1063/1.4958461} \BibitemShut {NoStop}%
\bibitem [{\citenamefont {Lietz}\ \emph {et~al.}(2017)\citenamefont {Lietz}, \citenamefont {Novario}, \citenamefont {Jansen}, \citenamefont {Hagen},\ and\ \citenamefont {Hjorth-Jensen}}]{Ref3}%
  \BibitemOpen
  \bibfield  {author} {\bibinfo {author} {\bibfnamefont {J.~G.}\ \bibnamefont {Lietz}}, \bibinfo {author} {\bibfnamefont {S.}~\bibnamefont {Novario}}, \bibinfo {author} {\bibfnamefont {G.~R.}\ \bibnamefont {Jansen}}, \bibinfo {author} {\bibfnamefont {G.}~\bibnamefont {Hagen}}, \ and\ \bibinfo {author} {\bibfnamefont {M.}~\bibnamefont {Hjorth-Jensen}},\ }\enquote {\bibinfo {title} {Computational nuclear physics and post hartree-fock methods},}\ in\ \href {\doibase 10.1007/978-3-319-53336-0_8} {\emph {\bibinfo {booktitle} {An Advanced Course in Computational Nuclear Physics: Bridging the Scales from Quarks to Neutron Stars}}},\ \bibinfo {editor} {edited by\ \bibinfo {editor} {\bibfnamefont {M.}~\bibnamefont {Hjorth-Jensen}}, \bibinfo {editor} {\bibfnamefont {M.~P.}\ \bibnamefont {Lombardo}}, \ and\ \bibinfo {editor} {\bibfnamefont {U.}~\bibnamefont {van Kolck}}}\ (\bibinfo  {publisher} {Springer International Publishing},\ \bibinfo {address} {Cham},\ \bibinfo {year} {2017})\ pp.\ \bibinfo {pages}
  {293--399}\BibitemShut {NoStop}%
\bibitem [{\citenamefont {Bishop}(1991)}]{Ref68}%
  \BibitemOpen
  \bibfield  {author} {\bibinfo {author} {\bibfnamefont {R.}~\bibnamefont {Bishop}},\ }\href {\doibase 10.1007/BF01119617} {\bibfield  {journal} {\bibinfo  {journal} {Theoretica Chimica Acta}\ }\textbf {\bibinfo {volume} {80}},\ \bibinfo {pages} {95} (\bibinfo {year} {1991})}\BibitemShut {NoStop}%
\bibitem [{\citenamefont {Bishop}\ and\ \citenamefont {L\"uhrmann}(1982)}]{Ref71}%
  \BibitemOpen
  \bibfield  {author} {\bibinfo {author} {\bibfnamefont {R.~F.}\ \bibnamefont {Bishop}}\ and\ \bibinfo {author} {\bibfnamefont {K.~H.}\ \bibnamefont {L\"uhrmann}},\ }\href {\doibase 10.1103/PhysRevB.26.5523} {\bibfield  {journal} {\bibinfo  {journal} {Phys. Rev. B}\ }\textbf {\bibinfo {volume} {26}},\ \bibinfo {pages} {5523} (\bibinfo {year} {1982})}\BibitemShut {NoStop}%
\bibitem [{\citenamefont {Bishop}\ and\ \citenamefont {Lührmann}(1981)}]{Ref72}%
  \BibitemOpen
  \bibfield  {author} {\bibinfo {author} {\bibfnamefont {R.}~\bibnamefont {Bishop}}\ and\ \bibinfo {author} {\bibfnamefont {K.}~\bibnamefont {Lührmann}},\ }\href {\doibase https://doi.org/10.1016/0378-4363(81)90741-5} {\bibfield  {journal} {\bibinfo  {journal} {Physica B+C}\ }\textbf {\bibinfo {volume} {108}},\ \bibinfo {pages} {873} (\bibinfo {year} {1981})}\BibitemShut {NoStop}%
\bibitem [{\citenamefont {Neufeld}\ and\ \citenamefont {Thom}(2017)}]{Ref95}%
  \BibitemOpen
  \bibfield  {author} {\bibinfo {author} {\bibfnamefont {V.~A.}\ \bibnamefont {Neufeld}}\ and\ \bibinfo {author} {\bibfnamefont {A.~J.~W.}\ \bibnamefont {Thom}},\ }\href {\doibase 10.1063/1.5003794} {\bibfield  {journal} {\bibinfo  {journal} {The Journal of Chemical Physics}\ }\textbf {\bibinfo {volume} {147}} (\bibinfo {year} {2017}),\ 10.1063/1.5003794},\ \bibinfo {note} {194105},\ \Eprint {http://arxiv.org/abs/https://pubs.aip.org/aip/jcp/article-pdf/doi/10.1063/1.5003794/15534755/194105\_1\_online.pdf} {https://pubs.aip.org/aip/jcp/article-pdf/doi/10.1063/1.5003794/15534755/194105\_1\_online.pdf} \BibitemShut {NoStop}%
\bibitem [{\citenamefont {Ceperley}\ and\ \citenamefont {Alder}(1980)}]{ceperley1980}%
  \BibitemOpen
  \bibfield  {author} {\bibinfo {author} {\bibfnamefont {D.~M.}\ \bibnamefont {Ceperley}}\ and\ \bibinfo {author} {\bibfnamefont {B.~J.}\ \bibnamefont {Alder}},\ }\href@noop {} {\bibfield  {journal} {\bibinfo  {journal} {Phys. Rev. Lett.}\ }\textbf {\bibinfo {volume} {45}},\ \bibinfo {pages} {566} (\bibinfo {year} {1980})}\BibitemShut {NoStop}%
\bibitem [{\citenamefont {Tanatar}\ and\ \citenamefont {Ceperley}(1989)}]{tanatar1989}%
  \BibitemOpen
  \bibfield  {author} {\bibinfo {author} {\bibfnamefont {B.}~\bibnamefont {Tanatar}}\ and\ \bibinfo {author} {\bibfnamefont {D.~M.}\ \bibnamefont {Ceperley}},\ }\href@noop {} {\bibfield  {journal} {\bibinfo  {journal} {Phys. Rev. B}\ }\textbf {\bibinfo {volume} {39}},\ \bibinfo {pages} {5005} (\bibinfo {year} {1989})}\BibitemShut {NoStop}%
\bibitem [{\citenamefont {Kwon}\ \emph {et~al.}(1998)\citenamefont {Kwon}, \citenamefont {Ceperley},\ and\ \citenamefont {Martin}}]{kwon1998}%
  \BibitemOpen
  \bibfield  {author} {\bibinfo {author} {\bibfnamefont {Y.}~\bibnamefont {Kwon}}, \bibinfo {author} {\bibfnamefont {D.~M.}\ \bibnamefont {Ceperley}}, \ and\ \bibinfo {author} {\bibfnamefont {R.~M.}\ \bibnamefont {Martin}},\ }\href@noop {} {\bibfield  {journal} {\bibinfo  {journal} {Phys. Rev. B}\ }\textbf {\bibinfo {volume} {58}},\ \bibinfo {pages} {6800} (\bibinfo {year} {1998})}\BibitemShut {NoStop}%
\bibitem [{\citenamefont {Pescia}\ \emph {et~al.}(2023)\citenamefont {Pescia}, \citenamefont {Nys}, \citenamefont {Kim}, \citenamefont {Lovato},\ and\ \citenamefont {Carleo}}]{kim2023}%
  \BibitemOpen
  \bibfield  {author} {\bibinfo {author} {\bibfnamefont {G.}~\bibnamefont {Pescia}}, \bibinfo {author} {\bibfnamefont {J.}~\bibnamefont {Nys}}, \bibinfo {author} {\bibfnamefont {J.}~\bibnamefont {Kim}}, \bibinfo {author} {\bibfnamefont {A.}~\bibnamefont {Lovato}}, \ and\ \bibinfo {author} {\bibfnamefont {G.}~\bibnamefont {Carleo}},\ }\href@noop {} {\enquote {\bibinfo {title} {Message-passing neural quantum states for the homogeneous electron gas},}\ } (\bibinfo {year} {2023}),\ \Eprint {http://arxiv.org/abs/2305.07240} {arXiv:2305.07240 [quant-ph]} \BibitemShut {NoStop}%
\bibitem [{\citenamefont {Cassella}\ \emph {et~al.}(2023)\citenamefont {Cassella}, \citenamefont {Sutterud}, \citenamefont {Azadi}, \citenamefont {Drummond}, \citenamefont {Pfau}, \citenamefont {Spencer},\ and\ \citenamefont {Foulkes}}]{cassella2023}%
  \BibitemOpen
  \bibfield  {author} {\bibinfo {author} {\bibfnamefont {G.}~\bibnamefont {Cassella}}, \bibinfo {author} {\bibfnamefont {H.}~\bibnamefont {Sutterud}}, \bibinfo {author} {\bibfnamefont {S.}~\bibnamefont {Azadi}}, \bibinfo {author} {\bibfnamefont {N.~D.}\ \bibnamefont {Drummond}}, \bibinfo {author} {\bibfnamefont {D.}~\bibnamefont {Pfau}}, \bibinfo {author} {\bibfnamefont {J.~S.}\ \bibnamefont {Spencer}}, \ and\ \bibinfo {author} {\bibfnamefont {W.~M.~C.}\ \bibnamefont {Foulkes}},\ }\href {\doibase 10.1103/PhysRevLett.130.036401} {\bibfield  {journal} {\bibinfo  {journal} {Phys. Rev. Lett.}\ }\textbf {\bibinfo {volume} {130}},\ \bibinfo {pages} {036401} (\bibinfo {year} {2023})}\BibitemShut {NoStop}%
\bibitem [{\citenamefont {Shavitt}\ and\ \citenamefont {Bartlett}(2009{\natexlab{a}})}]{Ref21}%
  \BibitemOpen
  \bibfield  {author} {\bibinfo {author} {\bibfnamefont {I.}~\bibnamefont {Shavitt}}\ and\ \bibinfo {author} {\bibfnamefont {R.~J.}\ \bibnamefont {Bartlett}},\ }\href {\doibase 10.1017/CBO9780511596834} {\emph {\bibinfo {title} {Many-Body Methods in Chemistry and Physics: MBPT and Coupled-Cluster Theory}}},\ Cambridge Molecular Science\ (\bibinfo  {publisher} {Cambridge University Press},\ \bibinfo {year} {2009})\BibitemShut {NoStop}%
\bibitem [{\citenamefont {Binder}\ \emph {et~al.}(2013)\citenamefont {Binder}, \citenamefont {Piecuch}, \citenamefont {Calci}, \citenamefont {Langhammer}, \citenamefont {Navr\'atil},\ and\ \citenamefont {Roth}}]{Ref47}%
  \BibitemOpen
  \bibfield  {author} {\bibinfo {author} {\bibfnamefont {S.}~\bibnamefont {Binder}}, \bibinfo {author} {\bibfnamefont {P.}~\bibnamefont {Piecuch}}, \bibinfo {author} {\bibfnamefont {A.}~\bibnamefont {Calci}}, \bibinfo {author} {\bibfnamefont {J.}~\bibnamefont {Langhammer}}, \bibinfo {author} {\bibfnamefont {P.}~\bibnamefont {Navr\'atil}}, \ and\ \bibinfo {author} {\bibfnamefont {R.}~\bibnamefont {Roth}},\ }\href {\doibase 10.1103/PhysRevC.88.054319} {\bibfield  {journal} {\bibinfo  {journal} {Phys. Rev. C}\ }\textbf {\bibinfo {volume} {88}},\ \bibinfo {pages} {054319} (\bibinfo {year} {2013})}\BibitemShut {NoStop}%
\bibitem [{\citenamefont {Bishop}\ and\ \citenamefont {Kümmel}(1987)}]{Ref70}%
  \BibitemOpen
  \bibfield  {author} {\bibinfo {author} {\bibfnamefont {R.~F.}\ \bibnamefont {Bishop}}\ and\ \bibinfo {author} {\bibfnamefont {H.~G.}\ \bibnamefont {Kümmel}},\ }\href {\doibase 10.1063/1.881103} {\bibfield  {journal} {\bibinfo  {journal} {Physics Today}\ }\textbf {\bibinfo {volume} {40}},\ \bibinfo {pages} {52} (\bibinfo {year} {1987})},\ \Eprint {http://arxiv.org/abs/https://pubs.aip.org/physicstoday/article-pdf/40/3/52/8298562/52\_1\_online.pdf} {https://pubs.aip.org/physicstoday/article-pdf/40/3/52/8298562/52\_1\_online.pdf} \BibitemShut {NoStop}%
\bibitem [{\citenamefont {Vargas-Hern\'andez}\ \emph {et~al.}(2018)\citenamefont {Vargas-Hern\'andez}, \citenamefont {Sous}, \citenamefont {Berciu},\ and\ \citenamefont {Krems}}]{Ref106}%
  \BibitemOpen
  \bibfield  {author} {\bibinfo {author} {\bibfnamefont {R.~A.}\ \bibnamefont {Vargas-Hern\'andez}}, \bibinfo {author} {\bibfnamefont {J.}~\bibnamefont {Sous}}, \bibinfo {author} {\bibfnamefont {M.}~\bibnamefont {Berciu}}, \ and\ \bibinfo {author} {\bibfnamefont {R.~V.}\ \bibnamefont {Krems}},\ }\href {\doibase 10.1103/PhysRevLett.121.255702} {\bibfield  {journal} {\bibinfo  {journal} {Phys. Rev. Lett.}\ }\textbf {\bibinfo {volume} {121}},\ \bibinfo {pages} {255702} (\bibinfo {year} {2018})}\BibitemShut {NoStop}%
\bibitem [{\citenamefont {Neufcourt}\ \emph {et~al.}(2018)\citenamefont {Neufcourt}, \citenamefont {Cao}, \citenamefont {Nazarewicz},\ and\ \citenamefont {Viens}}]{Ref107}%
  \BibitemOpen
  \bibfield  {author} {\bibinfo {author} {\bibfnamefont {L.}~\bibnamefont {Neufcourt}}, \bibinfo {author} {\bibfnamefont {Y.}~\bibnamefont {Cao}}, \bibinfo {author} {\bibfnamefont {W.}~\bibnamefont {Nazarewicz}}, \ and\ \bibinfo {author} {\bibfnamefont {F.}~\bibnamefont {Viens}},\ }\href {\doibase 10.1103/PhysRevC.98.034318} {\bibfield  {journal} {\bibinfo  {journal} {Phys. Rev. C}\ }\textbf {\bibinfo {volume} {98}},\ \bibinfo {pages} {034318} (\bibinfo {year} {2018})}\BibitemShut {NoStop}%
\bibitem [{\citenamefont {Kümmel}\ \emph {et~al.}(1978)\citenamefont {Kümmel}, \citenamefont {Lührmann},\ and\ \citenamefont {Zabolitzky}}]{Ref147}%
  \BibitemOpen
  \bibfield  {author} {\bibinfo {author} {\bibfnamefont {H.}~\bibnamefont {Kümmel}}, \bibinfo {author} {\bibfnamefont {K.}~\bibnamefont {Lührmann}}, \ and\ \bibinfo {author} {\bibfnamefont {J.}~\bibnamefont {Zabolitzky}},\ }\href {\doibase https://doi.org/10.1016/0370-1573(78)90081-9} {\bibfield  {journal} {\bibinfo  {journal} {Physics Reports}\ }\textbf {\bibinfo {volume} {36}},\ \bibinfo {pages} {1} (\bibinfo {year} {1978})}\BibitemShut {NoStop}%
\bibitem [{\citenamefont {Bartlett}\ and\ \citenamefont {Musia\l{}}(2007)}]{Ref149}%
  \BibitemOpen
  \bibfield  {author} {\bibinfo {author} {\bibfnamefont {R.~J.}\ \bibnamefont {Bartlett}}\ and\ \bibinfo {author} {\bibfnamefont {M.}~\bibnamefont {Musia\l{}}},\ }\href {\doibase 10.1103/RevModPhys.79.291} {\bibfield  {journal} {\bibinfo  {journal} {Rev. Mod. Phys.}\ }\textbf {\bibinfo {volume} {79}},\ \bibinfo {pages} {291} (\bibinfo {year} {2007})}\BibitemShut {NoStop}%
\bibitem [{\citenamefont {Čižek}\ and\ \citenamefont {Paldus}(1971)}]{Ref150}%
  \BibitemOpen
  \bibfield  {author} {\bibinfo {author} {\bibfnamefont {J.}~\bibnamefont {Čižek}}\ and\ \bibinfo {author} {\bibfnamefont {J.}~\bibnamefont {Paldus}},\ }\href {\doibase https://doi.org/10.1002/qua.560050402} {\bibfield  {journal} {\bibinfo  {journal} {International Journal of Quantum Chemistry}\ }\textbf {\bibinfo {volume} {5}},\ \bibinfo {pages} {359} (\bibinfo {year} {1971})},\ \Eprint {http://arxiv.org/abs/https://onlinelibrary.wiley.com/doi/pdf/10.1002/qua.560050402} {https://onlinelibrary.wiley.com/doi/pdf/10.1002/qua.560050402} \BibitemShut {NoStop}%
\bibitem [{\citenamefont {Čížek}(2004)}]{Ref151}%
  \BibitemOpen
  \bibfield  {author} {\bibinfo {author} {\bibfnamefont {J.}~\bibnamefont {Čížek}},\ }\href {\doibase 10.1063/1.1727484} {\bibfield  {journal} {\bibinfo  {journal} {The Journal of Chemical Physics}\ }\textbf {\bibinfo {volume} {45}},\ \bibinfo {pages} {4256} (\bibinfo {year} {2004})},\ \Eprint {http://arxiv.org/abs/https://pubs.aip.org/aip/jcp/article-pdf/45/11/4256/11266618/4256\_1\_online.pdf} {https://pubs.aip.org/aip/jcp/article-pdf/45/11/4256/11266618/4256\_1\_online.pdf} \BibitemShut {NoStop}%
\bibitem [{\citenamefont {Singal}\ and\ \citenamefont {Das}(1973)}]{singal1973}%
  \BibitemOpen
  \bibfield  {author} {\bibinfo {author} {\bibfnamefont {C.~M.}\ \bibnamefont {Singal}}\ and\ \bibinfo {author} {\bibfnamefont {T.~P.}\ \bibnamefont {Das}},\ }\href {\doibase 10.1103/PhysRevB.8.3675} {\bibfield  {journal} {\bibinfo  {journal} {Phys. Rev. B}\ }\textbf {\bibinfo {volume} {8}},\ \bibinfo {pages} {3675} (\bibinfo {year} {1973})}\BibitemShut {NoStop}%
\bibitem [{\citenamefont {Haftel}\ and\ \citenamefont {Tabakin}(1970)}]{haftel1970}%
  \BibitemOpen
  \bibfield  {author} {\bibinfo {author} {\bibfnamefont {M.~I.}\ \bibnamefont {Haftel}}\ and\ \bibinfo {author} {\bibfnamefont {F.}~\bibnamefont {Tabakin}},\ }\href@noop {} {\bibfield  {journal} {\bibinfo  {journal} {Nucl. Phys. A}\ }\textbf {\bibinfo {volume} {158}},\ \bibinfo {pages} {1} (\bibinfo {year} {1970})}\BibitemShut {NoStop}%
\bibitem [{\citenamefont {Freeman}(1977)}]{freeman1977}%
  \BibitemOpen
  \bibfield  {author} {\bibinfo {author} {\bibfnamefont {D.~L.}\ \bibnamefont {Freeman}},\ }\href@noop {} {\bibfield  {journal} {\bibinfo  {journal} {Phys. Rev. B}\ }\textbf {\bibinfo {volume} {15}},\ \bibinfo {pages} {5512} (\bibinfo {year} {1977})}\BibitemShut {NoStop}%
\bibitem [{\citenamefont {Bishop}\ and\ \citenamefont {L{"u}hrmann}(1978)}]{bishop1978}%
  \BibitemOpen
  \bibfield  {author} {\bibinfo {author} {\bibfnamefont {R.~F.}\ \bibnamefont {Bishop}}\ and\ \bibinfo {author} {\bibfnamefont {K.~H.}\ \bibnamefont {L{"u}hrmann}},\ }\href@noop {} {\bibfield  {journal} {\bibinfo  {journal} {Phys. Rev. B}\ }\textbf {\bibinfo {volume} {17}},\ \bibinfo {pages} {3757} (\bibinfo {year} {1978})}\BibitemShut {NoStop}%
\bibitem [{\citenamefont {Bishop}\ and\ \citenamefont {L{"u}hrmann}(1982)}]{bishop1982}%
  \BibitemOpen
  \bibfield  {author} {\bibinfo {author} {\bibfnamefont {R.~F.}\ \bibnamefont {Bishop}}\ and\ \bibinfo {author} {\bibfnamefont {K.~H.}\ \bibnamefont {L{"u}hrmann}},\ }\href@noop {} {\bibfield  {journal} {\bibinfo  {journal} {Phys. Rev. B}\ }\textbf {\bibinfo {volume} {26}},\ \bibinfo {pages} {5523} (\bibinfo {year} {1982})}\BibitemShut {NoStop}%
\bibitem [{\citenamefont {Shavitt}\ and\ \citenamefont {Bartlett}(2009{\natexlab{b}})}]{shavittbartlett2009}%
  \BibitemOpen
  \bibfield  {author} {\bibinfo {author} {\bibfnamefont {I.}~\bibnamefont {Shavitt}}\ and\ \bibinfo {author} {\bibfnamefont {R.~J.}\ \bibnamefont {Bartlett}},\ }\href@noop {} {\emph {\bibinfo {title} {Many-body Methods in Chemistry and Physics}}}\ (\bibinfo  {publisher} {Cambridge University Press},\ \bibinfo {year} {2009})\BibitemShut {NoStop}%
\bibitem [{\citenamefont {Shepherd}\ \emph {et~al.}(2012{\natexlab{b}})\citenamefont {Shepherd}, \citenamefont {Gr\"uneis}, \citenamefont {Booth}, \citenamefont {Kresse},\ and\ \citenamefont {Alavi}}]{shepherd2012b}%
  \BibitemOpen
  \bibfield  {author} {\bibinfo {author} {\bibfnamefont {J.~J.}\ \bibnamefont {Shepherd}}, \bibinfo {author} {\bibfnamefont {A.}~\bibnamefont {Gr\"uneis}}, \bibinfo {author} {\bibfnamefont {G.~H.}\ \bibnamefont {Booth}}, \bibinfo {author} {\bibfnamefont {G.}~\bibnamefont {Kresse}}, \ and\ \bibinfo {author} {\bibfnamefont {A.}~\bibnamefont {Alavi}},\ }\href@noop {} {\bibfield  {journal} {\bibinfo  {journal} {Phys. Rev. B}\ }\textbf {\bibinfo {volume} {86}},\ \bibinfo {pages} {035111} (\bibinfo {year} {2012}{\natexlab{b}})}\BibitemShut {NoStop}%
\bibitem [{\citenamefont {Shepherd}\ and\ \citenamefont {Gr\"uneis}(2013)}]{shepherd2013a}%
  \BibitemOpen
  \bibfield  {author} {\bibinfo {author} {\bibfnamefont {J.~J.}\ \bibnamefont {Shepherd}}\ and\ \bibinfo {author} {\bibfnamefont {A.}~\bibnamefont {Gr\"uneis}},\ }\href {\doibase 10.1103/PhysRevLett.110.226401} {\bibfield  {journal} {\bibinfo  {journal} {Phys. Rev. Lett.}\ }\textbf {\bibinfo {volume} {110}},\ \bibinfo {pages} {226401} (\bibinfo {year} {2013})}\BibitemShut {NoStop}%
\bibitem [{\citenamefont {Shepherd}\ \emph {et~al.}(2014{\natexlab{a}})\citenamefont {Shepherd}, \citenamefont {Henderson},\ and\ \citenamefont {Scuseria}}]{shepherd2013b}%
  \BibitemOpen
  \bibfield  {author} {\bibinfo {author} {\bibfnamefont {J.~J.}\ \bibnamefont {Shepherd}}, \bibinfo {author} {\bibfnamefont {T.~M.}\ \bibnamefont {Henderson}}, \ and\ \bibinfo {author} {\bibfnamefont {G.~E.}\ \bibnamefont {Scuseria}},\ }\href {\doibase 10.1103/PhysRevLett.112.209901} {\bibfield  {journal} {\bibinfo  {journal} {Phys. Rev. Lett.}\ }\textbf {\bibinfo {volume} {112}},\ \bibinfo {pages} {209901} (\bibinfo {year} {2014}{\natexlab{a}})}\BibitemShut {NoStop}%
\bibitem [{\citenamefont {Shepherd}\ \emph {et~al.}(2014{\natexlab{b}})\citenamefont {Shepherd}, \citenamefont {Henderson},\ and\ \citenamefont {Scuseria}}]{shepherd2013c}%
  \BibitemOpen
  \bibfield  {author} {\bibinfo {author} {\bibfnamefont {J.~J.}\ \bibnamefont {Shepherd}}, \bibinfo {author} {\bibfnamefont {T.~M.}\ \bibnamefont {Henderson}}, \ and\ \bibinfo {author} {\bibfnamefont {G.~E.}\ \bibnamefont {Scuseria}},\ }\href {\doibase http://dx.doi.org/10.1063/1.4867783} {\bibfield  {journal} {\bibinfo  {journal} {The Journal of Chemical Physics}\ }\textbf {\bibinfo {volume} {140}},\ \bibinfo {eid} {124102} (\bibinfo {year} {2014}{\natexlab{b}}),\ http://dx.doi.org/10.1063/1.4867783}\BibitemShut {NoStop}%
\bibitem [{\citenamefont {Murphy}(2013)}]{Ref11}%
  \BibitemOpen
  \bibfield  {author} {\bibinfo {author} {\bibfnamefont {K.~P.}\ \bibnamefont {Murphy}},\ }\href {https://www.amazon.com/Machine-Learning-Probabilistic-Perspective-Computation/dp/0262018020/ref=sr_1_2?ie=UTF8&qid=1336857747&sr=8-2} {\emph {\bibinfo {title} {Machine learning : a probabilistic perspective}}}\ (\bibinfo  {publisher} {MIT Press},\ \bibinfo {address} {Cambridge, Mass. [u.a.]},\ \bibinfo {year} {2013})\BibitemShut {NoStop}%
\bibitem [{\citenamefont {Géron}(2017)}]{Ref12}%
  \BibitemOpen
  \bibfield  {author} {\bibinfo {author} {\bibfnamefont {A.}~\bibnamefont {Géron}},\ }\href@noop {} {\emph {\bibinfo {title} {Hands-on machine learning with Scikit-Learn and TensorFlow : concepts, tools, and techniques to build intelligent systems}}}\ (\bibinfo  {publisher} {O'Reilly Media},\ \bibinfo {address} {Sebastopol, CA},\ \bibinfo {year} {2017})\BibitemShut {NoStop}%
\bibitem [{\citenamefont {Mehta}\ \emph {et~al.}(2019)\citenamefont {Mehta}, \citenamefont {Bukov}, \citenamefont {Wang}, \citenamefont {Day}, \citenamefont {Richardson}, \citenamefont {Fisher},\ and\ \citenamefont {Schwab}}]{Ref19}%
  \BibitemOpen
  \bibfield  {author} {\bibinfo {author} {\bibfnamefont {P.}~\bibnamefont {Mehta}}, \bibinfo {author} {\bibfnamefont {M.}~\bibnamefont {Bukov}}, \bibinfo {author} {\bibfnamefont {C.-H.}\ \bibnamefont {Wang}}, \bibinfo {author} {\bibfnamefont {A.~G.}\ \bibnamefont {Day}}, \bibinfo {author} {\bibfnamefont {C.}~\bibnamefont {Richardson}}, \bibinfo {author} {\bibfnamefont {C.~K.}\ \bibnamefont {Fisher}}, \ and\ \bibinfo {author} {\bibfnamefont {D.~J.}\ \bibnamefont {Schwab}},\ }\href {\doibase https://doi.org/10.1016/j.physrep.2019.03.001} {\bibfield  {journal} {\bibinfo  {journal} {Physics Reports}\ }\textbf {\bibinfo {volume} {810}},\ \bibinfo {pages} {1} (\bibinfo {year} {2019})},\ \bibinfo {note} {a high-bias, low-variance introduction to Machine Learning for physicists}\BibitemShut {NoStop}%
\bibitem [{\citenamefont {Negoita}\ \emph {et~al.}(2019)\citenamefont {Negoita}, \citenamefont {Vary}, \citenamefont {Luecke}, \citenamefont {Maris}, \citenamefont {Shirokov}, \citenamefont {Shin}, \citenamefont {Kim}, \citenamefont {Ng}, \citenamefont {Yang}, \citenamefont {Lockner},\ and\ \citenamefont {Prabhu}}]{Ref23}%
  \BibitemOpen
  \bibfield  {author} {\bibinfo {author} {\bibfnamefont {G.~A.}\ \bibnamefont {Negoita}}, \bibinfo {author} {\bibfnamefont {J.~P.}\ \bibnamefont {Vary}}, \bibinfo {author} {\bibfnamefont {G.~R.}\ \bibnamefont {Luecke}}, \bibinfo {author} {\bibfnamefont {P.}~\bibnamefont {Maris}}, \bibinfo {author} {\bibfnamefont {A.~M.}\ \bibnamefont {Shirokov}}, \bibinfo {author} {\bibfnamefont {I.~J.}\ \bibnamefont {Shin}}, \bibinfo {author} {\bibfnamefont {Y.}~\bibnamefont {Kim}}, \bibinfo {author} {\bibfnamefont {E.~G.}\ \bibnamefont {Ng}}, \bibinfo {author} {\bibfnamefont {C.}~\bibnamefont {Yang}}, \bibinfo {author} {\bibfnamefont {M.}~\bibnamefont {Lockner}}, \ and\ \bibinfo {author} {\bibfnamefont {G.~M.}\ \bibnamefont {Prabhu}},\ }\href {\doibase 10.1103/PhysRevC.99.054308} {\bibfield  {journal} {\bibinfo  {journal} {Phys. Rev. C}\ }\textbf {\bibinfo {volume} {99}},\ \bibinfo {pages} {054308} (\bibinfo {year} {2019})}\BibitemShut {NoStop}%
\bibitem [{\citenamefont {Jiang}\ \emph {et~al.}(2019)\citenamefont {Jiang}, \citenamefont {Hagen},\ and\ \citenamefont {Papenbrock}}]{Ref6}%
  \BibitemOpen
  \bibfield  {author} {\bibinfo {author} {\bibfnamefont {W.~G.}\ \bibnamefont {Jiang}}, \bibinfo {author} {\bibfnamefont {G.}~\bibnamefont {Hagen}}, \ and\ \bibinfo {author} {\bibfnamefont {T.}~\bibnamefont {Papenbrock}},\ }\href {\doibase 10.1103/PhysRevC.100.054326} {\bibfield  {journal} {\bibinfo  {journal} {Phys. Rev. C}\ }\textbf {\bibinfo {volume} {100}},\ \bibinfo {pages} {054326} (\bibinfo {year} {2019})}\BibitemShut {NoStop}%
\bibitem [{\citenamefont {Akkoyun}\ \emph {et~al.}(2013)\citenamefont {Akkoyun}, \citenamefont {Bayram}, \citenamefont {Kara},\ and\ \citenamefont {Sinan}}]{Ref27}%
  \BibitemOpen
  \bibfield  {author} {\bibinfo {author} {\bibfnamefont {S.}~\bibnamefont {Akkoyun}}, \bibinfo {author} {\bibfnamefont {T.}~\bibnamefont {Bayram}}, \bibinfo {author} {\bibfnamefont {S.~O.}\ \bibnamefont {Kara}}, \ and\ \bibinfo {author} {\bibfnamefont {A.}~\bibnamefont {Sinan}},\ }\href {\doibase 10.1088/0954-3899/40/5/055106} {\bibfield  {journal} {\bibinfo  {journal} {Journal of Physics G: Nuclear and Particle Physics}\ }\textbf {\bibinfo {volume} {40}},\ \bibinfo {pages} {055106} (\bibinfo {year} {2013})}\BibitemShut {NoStop}%
\bibitem [{\citenamefont {Utama}\ \emph {et~al.}(2016)\citenamefont {Utama}, \citenamefont {Piekarewicz},\ and\ \citenamefont {Prosper}}]{Ref28}%
  \BibitemOpen
  \bibfield  {author} {\bibinfo {author} {\bibfnamefont {R.}~\bibnamefont {Utama}}, \bibinfo {author} {\bibfnamefont {J.}~\bibnamefont {Piekarewicz}}, \ and\ \bibinfo {author} {\bibfnamefont {H.~B.}\ \bibnamefont {Prosper}},\ }\href {\doibase 10.1103/PhysRevC.93.014311} {\bibfield  {journal} {\bibinfo  {journal} {Phys. Rev. C}\ }\textbf {\bibinfo {volume} {93}},\ \bibinfo {pages} {014311} (\bibinfo {year} {2016})}\BibitemShut {NoStop}%
\bibitem [{\citenamefont {Carleo}\ and\ \citenamefont {Troyer}(2017)}]{Ref32}%
  \BibitemOpen
  \bibfield  {author} {\bibinfo {author} {\bibfnamefont {G.}~\bibnamefont {Carleo}}\ and\ \bibinfo {author} {\bibfnamefont {M.}~\bibnamefont {Troyer}},\ }\href {\doibase 10.1126/science.aag2302} {\bibfield  {journal} {\bibinfo  {journal} {Science}\ }\textbf {\bibinfo {volume} {355}},\ \bibinfo {pages} {602} (\bibinfo {year} {2017})},\ \Eprint {http://arxiv.org/abs/https://www.science.org/doi/pdf/10.1126/science.aag2302} {https://www.science.org/doi/pdf/10.1126/science.aag2302} \BibitemShut {NoStop}%
\bibitem [{\citenamefont {Boehnlein}\ \emph {et~al.}(2022)\citenamefont {Boehnlein}, \citenamefont {Diefenthaler}, \citenamefont {Sato}, \citenamefont {Schram}, \citenamefont {Ziegler}, \citenamefont {Fanelli}, \citenamefont {Hjorth-Jensen}, \citenamefont {Horn}, \citenamefont {Kuchera}, \citenamefont {Lee}, \citenamefont {Nazarewicz}, \citenamefont {Ostroumov}, \citenamefont {Orginos}, \citenamefont {Poon}, \citenamefont {Wang}, \citenamefont {Scheinker}, \citenamefont {Smith},\ and\ \citenamefont {Pang}}]{Ref109}%
  \BibitemOpen
  \bibfield  {author} {\bibinfo {author} {\bibfnamefont {A.}~\bibnamefont {Boehnlein}}, \bibinfo {author} {\bibfnamefont {M.}~\bibnamefont {Diefenthaler}}, \bibinfo {author} {\bibfnamefont {N.}~\bibnamefont {Sato}}, \bibinfo {author} {\bibfnamefont {M.}~\bibnamefont {Schram}}, \bibinfo {author} {\bibfnamefont {V.}~\bibnamefont {Ziegler}}, \bibinfo {author} {\bibfnamefont {C.}~\bibnamefont {Fanelli}}, \bibinfo {author} {\bibfnamefont {M.}~\bibnamefont {Hjorth-Jensen}}, \bibinfo {author} {\bibfnamefont {T.}~\bibnamefont {Horn}}, \bibinfo {author} {\bibfnamefont {M.~P.}\ \bibnamefont {Kuchera}}, \bibinfo {author} {\bibfnamefont {D.}~\bibnamefont {Lee}}, \bibinfo {author} {\bibfnamefont {W.}~\bibnamefont {Nazarewicz}}, \bibinfo {author} {\bibfnamefont {P.}~\bibnamefont {Ostroumov}}, \bibinfo {author} {\bibfnamefont {K.}~\bibnamefont {Orginos}}, \bibinfo {author} {\bibfnamefont {A.}~\bibnamefont {Poon}}, \bibinfo {author} {\bibfnamefont {X.-N.}\ \bibnamefont {Wang}}, \bibinfo {author} {\bibfnamefont
  {A.}~\bibnamefont {Scheinker}}, \bibinfo {author} {\bibfnamefont {M.~S.}\ \bibnamefont {Smith}}, \ and\ \bibinfo {author} {\bibfnamefont {L.-G.}\ \bibnamefont {Pang}},\ }\href {\doibase 10.1103/RevModPhys.94.031003} {\bibfield  {journal} {\bibinfo  {journal} {Rev. Mod. Phys.}\ }\textbf {\bibinfo {volume} {94}},\ \bibinfo {pages} {031003} (\bibinfo {year} {2022})}\BibitemShut {NoStop}%
\bibitem [{\citenamefont {Ismail}\ and\ \citenamefont {Gezerlis}(2021)}]{Ref110}%
  \BibitemOpen
  \bibfield  {author} {\bibinfo {author} {\bibfnamefont {N.}~\bibnamefont {Ismail}}\ and\ \bibinfo {author} {\bibfnamefont {A.}~\bibnamefont {Gezerlis}},\ }\href {\doibase 10.1103/PhysRevC.104.055802} {\bibfield  {journal} {\bibinfo  {journal} {Phys. Rev. C}\ }\textbf {\bibinfo {volume} {104}},\ \bibinfo {pages} {055802} (\bibinfo {year} {2021})}\BibitemShut {NoStop}%
\bibitem [{\citenamefont {Meredig}\ \emph {et~al.}(2018)\citenamefont {Meredig}, \citenamefont {Antono}, \citenamefont {Church}, \citenamefont {Hutchinson}, \citenamefont {Ling}, \citenamefont {Paradiso}, \citenamefont {Blaiszik}, \citenamefont {Foster}, \citenamefont {Gibbons}, \citenamefont {Hattrick-Simpers}, \citenamefont {Mehta},\ and\ \citenamefont {Ward}}]{Ref112}%
  \BibitemOpen
  \bibfield  {author} {\bibinfo {author} {\bibfnamefont {B.}~\bibnamefont {Meredig}}, \bibinfo {author} {\bibfnamefont {E.}~\bibnamefont {Antono}}, \bibinfo {author} {\bibfnamefont {C.}~\bibnamefont {Church}}, \bibinfo {author} {\bibfnamefont {M.}~\bibnamefont {Hutchinson}}, \bibinfo {author} {\bibfnamefont {J.}~\bibnamefont {Ling}}, \bibinfo {author} {\bibfnamefont {S.}~\bibnamefont {Paradiso}}, \bibinfo {author} {\bibfnamefont {B.}~\bibnamefont {Blaiszik}}, \bibinfo {author} {\bibfnamefont {I.}~\bibnamefont {Foster}}, \bibinfo {author} {\bibfnamefont {B.}~\bibnamefont {Gibbons}}, \bibinfo {author} {\bibfnamefont {J.}~\bibnamefont {Hattrick-Simpers}}, \bibinfo {author} {\bibfnamefont {A.}~\bibnamefont {Mehta}}, \ and\ \bibinfo {author} {\bibfnamefont {L.}~\bibnamefont {Ward}},\ }\href {\doibase 10.1039/C8ME00012C} {\bibfield  {journal} {\bibinfo  {journal} {Mol. Syst. Des. Eng.}\ }\textbf {\bibinfo {volume} {3}},\ \bibinfo {pages} {819} (\bibinfo {year} {2018})}\BibitemShut {NoStop}%
\bibitem [{\citenamefont {Jinnouchi}\ \emph {et~al.}(2017)\citenamefont {Jinnouchi}, \citenamefont {Hirata},\ and\ \citenamefont {Asahi}}]{Ref114}%
  \BibitemOpen
  \bibfield  {author} {\bibinfo {author} {\bibfnamefont {R.}~\bibnamefont {Jinnouchi}}, \bibinfo {author} {\bibfnamefont {H.}~\bibnamefont {Hirata}}, \ and\ \bibinfo {author} {\bibfnamefont {R.}~\bibnamefont {Asahi}},\ }\href {\doibase 10.1021/acs.jpcc.7b08686} {\bibfield  {journal} {\bibinfo  {journal} {The Journal of Physical Chemistry C}\ }\textbf {\bibinfo {volume} {121}},\ \bibinfo {pages} {26397} (\bibinfo {year} {2017})},\ \Eprint {http://arxiv.org/abs/https://doi.org/10.1021/acs.jpcc.7b08686} {https://doi.org/10.1021/acs.jpcc.7b08686} \BibitemShut {NoStop}%
\bibitem [{\citenamefont {Knöll}\ \emph {et~al.}(2023)\citenamefont {Knöll}, \citenamefont {Wolfgruber}, \citenamefont {Agel}, \citenamefont {Wenz},\ and\ \citenamefont {Roth}}]{Ref117}%
  \BibitemOpen
  \bibfield  {author} {\bibinfo {author} {\bibfnamefont {M.}~\bibnamefont {Knöll}}, \bibinfo {author} {\bibfnamefont {T.}~\bibnamefont {Wolfgruber}}, \bibinfo {author} {\bibfnamefont {M.~L.}\ \bibnamefont {Agel}}, \bibinfo {author} {\bibfnamefont {C.}~\bibnamefont {Wenz}}, \ and\ \bibinfo {author} {\bibfnamefont {R.}~\bibnamefont {Roth}},\ }\href {\doibase https://doi.org/10.1016/j.physletb.2023.137781} {\bibfield  {journal} {\bibinfo  {journal} {Physics Letters B}\ }\textbf {\bibinfo {volume} {839}},\ \bibinfo {pages} {137781} (\bibinfo {year} {2023})}\BibitemShut {NoStop}%
\bibitem [{\citenamefont {Margraf}\ and\ \citenamefont {Reuter}(2018)}]{Ref7}%
  \BibitemOpen
  \bibfield  {author} {\bibinfo {author} {\bibfnamefont {J.~T.}\ \bibnamefont {Margraf}}\ and\ \bibinfo {author} {\bibfnamefont {K.}~\bibnamefont {Reuter}},\ }\href {\doibase 10.1021/acs.jpca.8b04455} {\bibfield  {journal} {\bibinfo  {journal} {The Journal of Physical Chemistry A}\ }\textbf {\bibinfo {volume} {122}},\ \bibinfo {pages} {6343} (\bibinfo {year} {2018})},\ \bibinfo {note} {pMID: 29985611},\ \Eprint {http://arxiv.org/abs/https://doi.org/10.1021/acs.jpca.8b04455} {https://doi.org/10.1021/acs.jpca.8b04455} \BibitemShut {NoStop}%
\bibitem [{\citenamefont {Smith}\ \emph {et~al.}(2019)\citenamefont {Smith}, \citenamefont {Nebgen}, \citenamefont {Zubatyuk}, \citenamefont {Lubbers}, \citenamefont {Devereux}, \citenamefont {Barros}, \citenamefont {Tretiak}, \citenamefont {Isayev},\ and\ \citenamefont {Roitberg}}]{Ref81}%
  \BibitemOpen
  \bibfield  {author} {\bibinfo {author} {\bibfnamefont {J.}~\bibnamefont {Smith}}, \bibinfo {author} {\bibfnamefont {B.}~\bibnamefont {Nebgen}}, \bibinfo {author} {\bibfnamefont {R.}~\bibnamefont {Zubatyuk}}, \bibinfo {author} {\bibfnamefont {N.}~\bibnamefont {Lubbers}}, \bibinfo {author} {\bibfnamefont {C.}~\bibnamefont {Devereux}}, \bibinfo {author} {\bibfnamefont {K.}~\bibnamefont {Barros}}, \bibinfo {author} {\bibfnamefont {S.}~\bibnamefont {Tretiak}}, \bibinfo {author} {\bibfnamefont {O.}~\bibnamefont {Isayev}}, \ and\ \bibinfo {author} {\bibfnamefont {A.}~\bibnamefont {Roitberg}},\ }\href {\doibase 10.1038/s41467-019-10827-4} {\bibfield  {journal} {\bibinfo  {journal} {Nature Communications}\ }\textbf {\bibinfo {volume} {10}} (\bibinfo {year} {2019}),\ 10.1038/s41467-019-10827-4}\BibitemShut {NoStop}%
\bibitem [{\citenamefont {Mihm}\ \emph {et~al.}(2021{\natexlab{a}})\citenamefont {Mihm}, \citenamefont {Yang},\ and\ \citenamefont {Shepherd}}]{Ref74}%
  \BibitemOpen
  \bibfield  {author} {\bibinfo {author} {\bibfnamefont {T.~N.}\ \bibnamefont {Mihm}}, \bibinfo {author} {\bibfnamefont {B.}~\bibnamefont {Yang}}, \ and\ \bibinfo {author} {\bibfnamefont {J.~J.}\ \bibnamefont {Shepherd}},\ }\href {\doibase 10.1021/acs.jctc.0c01171} {\bibfield  {journal} {\bibinfo  {journal} {Journal of Chemical Theory and Computation}\ }\textbf {\bibinfo {volume} {17}},\ \bibinfo {pages} {2752} (\bibinfo {year} {2021}{\natexlab{a}})},\ \bibinfo {note} {pMID: 33830754},\ \Eprint {http://arxiv.org/abs/https://doi.org/10.1021/acs.jctc.0c01171} {https://doi.org/10.1021/acs.jctc.0c01171} \BibitemShut {NoStop}%
\bibitem [{\citenamefont {Mihm}\ \emph {et~al.}(2023)\citenamefont {Mihm}, \citenamefont {Weiler},\ and\ \citenamefont {Shepherd}}]{Ref86}%
  \BibitemOpen
  \bibfield  {author} {\bibinfo {author} {\bibfnamefont {T.~N.}\ \bibnamefont {Mihm}}, \bibinfo {author} {\bibfnamefont {L.}~\bibnamefont {Weiler}}, \ and\ \bibinfo {author} {\bibfnamefont {J.~J.}\ \bibnamefont {Shepherd}},\ }\href {\doibase 10.1021/acs.jctc.2c00737} {\bibfield  {journal} {\bibinfo  {journal} {Journal of Chemical Theory and Computation}\ }\textbf {\bibinfo {volume} {19}},\ \bibinfo {pages} {1686} (\bibinfo {year} {2023})},\ \bibinfo {note} {pMID: 36918372},\ \Eprint {http://arxiv.org/abs/https://doi.org/10.1021/acs.jctc.2c00737} {https://doi.org/10.1021/acs.jctc.2c00737} \BibitemShut {NoStop}%
\bibitem [{\citenamefont {Mihm}\ \emph {et~al.}(2021{\natexlab{b}})\citenamefont {Mihm}, \citenamefont {Van~Benschoten},\ and\ \citenamefont {Shepherd}}]{Ref94}%
  \BibitemOpen
  \bibfield  {author} {\bibinfo {author} {\bibfnamefont {T.~N.}\ \bibnamefont {Mihm}}, \bibinfo {author} {\bibfnamefont {W.~Z.}\ \bibnamefont {Van~Benschoten}}, \ and\ \bibinfo {author} {\bibfnamefont {J.~J.}\ \bibnamefont {Shepherd}},\ }\href {\doibase 10.1063/5.0033408} {\bibfield  {journal} {\bibinfo  {journal} {The Journal of Chemical Physics}\ }\textbf {\bibinfo {volume} {154}} (\bibinfo {year} {2021}{\natexlab{b}}),\ 10.1063/5.0033408},\ \bibinfo {note} {024113},\ \Eprint {http://arxiv.org/abs/https://pubs.aip.org/aip/jcp/article-pdf/doi/10.1063/5.0033408/15583886/024113\_1\_online.pdf} {https://pubs.aip.org/aip/jcp/article-pdf/doi/10.1063/5.0033408/15583886/024113\_1\_online.pdf} \BibitemShut {NoStop}%
\bibitem [{\citenamefont {Liao}\ and\ \citenamefont {Grüneis}(2016)}]{Ref96}%
  \BibitemOpen
  \bibfield  {author} {\bibinfo {author} {\bibfnamefont {K.}~\bibnamefont {Liao}}\ and\ \bibinfo {author} {\bibfnamefont {A.}~\bibnamefont {Grüneis}},\ }\href {\doibase 10.1063/1.4964307} {\bibfield  {journal} {\bibinfo  {journal} {The Journal of Chemical Physics}\ }\textbf {\bibinfo {volume} {145}} (\bibinfo {year} {2016}),\ 10.1063/1.4964307},\ \bibinfo {note} {141102},\ \Eprint {http://arxiv.org/abs/https://pubs.aip.org/aip/jcp/article-pdf/doi/10.1063/1.4964307/15518055/141102\_1\_online.pdf} {https://pubs.aip.org/aip/jcp/article-pdf/doi/10.1063/1.4964307/15518055/141102\_1\_online.pdf} \BibitemShut {NoStop}%
\bibitem [{\citenamefont {Mihm}\ \emph {et~al.}(2021{\natexlab{c}})\citenamefont {Mihm}, \citenamefont {Schäfer},\ and\ \citenamefont {Ramadugu}}]{Ref98}%
  \BibitemOpen
  \bibfield  {author} {\bibinfo {author} {\bibfnamefont {T.}~\bibnamefont {Mihm}}, \bibinfo {author} {\bibfnamefont {T.}~\bibnamefont {Schäfer}}, \ and\ \bibinfo {author} {\bibfnamefont {S.~e.~a.}\ \bibnamefont {Ramadugu}},\ }\href {\doibase 10.1038/s43588-021-00165-1} {\bibfield  {journal} {\bibinfo  {journal} {Nat Comput Sci}\ }\textbf {\bibinfo {volume} {1}},\ \bibinfo {pages} {801} (\bibinfo {year} {2021}{\natexlab{c}})}\BibitemShut {NoStop}%
\bibitem [{\citenamefont {Köhn}\ and\ \citenamefont {Tew}(2010)}]{Ref99}%
  \BibitemOpen
  \bibfield  {author} {\bibinfo {author} {\bibfnamefont {A.}~\bibnamefont {Köhn}}\ and\ \bibinfo {author} {\bibfnamefont {D.~P.}\ \bibnamefont {Tew}},\ }\href {\doibase 10.1063/1.3291040} {\bibfield  {journal} {\bibinfo  {journal} {The Journal of Chemical Physics}\ }\textbf {\bibinfo {volume} {132}} (\bibinfo {year} {2010}),\ 10.1063/1.3291040},\ \bibinfo {note} {024101},\ \Eprint {http://arxiv.org/abs/https://pubs.aip.org/aip/jcp/article-pdf/doi/10.1063/1.3291040/14745329/024101\_1\_online.pdf} {https://pubs.aip.org/aip/jcp/article-pdf/doi/10.1063/1.3291040/14745329/024101\_1\_online.pdf} \BibitemShut {NoStop}%
\bibitem [{\citenamefont {Coester}\ and\ \citenamefont {Kümmel}(1960)}]{Ref152}%
  \BibitemOpen
  \bibfield  {author} {\bibinfo {author} {\bibfnamefont {F.}~\bibnamefont {Coester}}\ and\ \bibinfo {author} {\bibfnamefont {H.}~\bibnamefont {Kümmel}},\ }\href {\doibase https://doi.org/10.1016/0029-5582(60)90140-1} {\bibfield  {journal} {\bibinfo  {journal} {Nuclear Physics}\ }\textbf {\bibinfo {volume} {17}},\ \bibinfo {pages} {477} (\bibinfo {year} {1960})}\BibitemShut {NoStop}%
\bibitem [{\citenamefont {Coester}(1958)}]{Ref153}%
  \BibitemOpen
  \bibfield  {author} {\bibinfo {author} {\bibfnamefont {F.}~\bibnamefont {Coester}},\ }\href {\doibase https://doi.org/10.1016/0029-5582(58)90280-3} {\bibfield  {journal} {\bibinfo  {journal} {Nuclear Physics}\ }\textbf {\bibinfo {volume} {7}},\ \bibinfo {pages} {421} (\bibinfo {year} {1958})}\BibitemShut {NoStop}%
\bibitem [{\citenamefont {Urban}\ \emph {et~al.}(1985)\citenamefont {Urban}, \citenamefont {Noga}, \citenamefont {Cole},\ and\ \citenamefont {Bartlett}}]{Ref157}%
  \BibitemOpen
  \bibfield  {author} {\bibinfo {author} {\bibfnamefont {M.}~\bibnamefont {Urban}}, \bibinfo {author} {\bibfnamefont {J.}~\bibnamefont {Noga}}, \bibinfo {author} {\bibfnamefont {S.~J.}\ \bibnamefont {Cole}}, \ and\ \bibinfo {author} {\bibfnamefont {R.~J.}\ \bibnamefont {Bartlett}},\ }\href {\doibase 10.1063/1.449067} {\bibfield  {journal} {\bibinfo  {journal} {The Journal of Chemical Physics}\ }\textbf {\bibinfo {volume} {83}},\ \bibinfo {pages} {4041–4046} (\bibinfo {year} {1985})}\BibitemShut {NoStop}%
\bibitem [{\citenamefont {Taube}\ and\ \citenamefont {Bartlett}(2008)}]{Ref140}%
  \BibitemOpen
  \bibfield  {author} {\bibinfo {author} {\bibfnamefont {A.~G.}\ \bibnamefont {Taube}}\ and\ \bibinfo {author} {\bibfnamefont {R.~J.}\ \bibnamefont {Bartlett}},\ }\href {\doibase 10.1063/1.2830236} {\bibfield  {journal} {\bibinfo  {journal} {The Journal of Chemical Physics}\ }\textbf {\bibinfo {volume} {128}} (\bibinfo {year} {2008}),\ 10.1063/1.2830236},\ \bibinfo {note} {044110},\ \Eprint {http://arxiv.org/abs/https://pubs.aip.org/aip/jcp/article-pdf/doi/10.1063/1.2830236/13473501/044110\_1\_online.pdf} {https://pubs.aip.org/aip/jcp/article-pdf/doi/10.1063/1.2830236/13473501/044110\_1\_online.pdf} \BibitemShut {NoStop}%
\bibitem [{\citenamefont {Kucharski}\ and\ \citenamefont {Bartlett}(1998)}]{Ref141}%
  \BibitemOpen
  \bibfield  {author} {\bibinfo {author} {\bibfnamefont {S.~A.}\ \bibnamefont {Kucharski}}\ and\ \bibinfo {author} {\bibfnamefont {R.~J.}\ \bibnamefont {Bartlett}},\ }\href {\doibase 10.1063/1.475961} {\bibfield  {journal} {\bibinfo  {journal} {The Journal of Chemical Physics}\ }\textbf {\bibinfo {volume} {108}},\ \bibinfo {pages} {5243} (\bibinfo {year} {1998})},\ \Eprint {http://arxiv.org/abs/https://pubs.aip.org/aip/jcp/article-pdf/108/13/5243/10789764/5243\_1\_online.pdf} {https://pubs.aip.org/aip/jcp/article-pdf/108/13/5243/10789764/5243\_1\_online.pdf} \BibitemShut {NoStop}%
\bibitem [{\citenamefont {Bartlett}\ \emph {et~al.}(1990)\citenamefont {Bartlett}, \citenamefont {Watts}, \citenamefont {Kucharski},\ and\ \citenamefont {Noga}}]{Ref142}%
  \BibitemOpen
  \bibfield  {author} {\bibinfo {author} {\bibfnamefont {R.~J.}\ \bibnamefont {Bartlett}}, \bibinfo {author} {\bibfnamefont {J.}~\bibnamefont {Watts}}, \bibinfo {author} {\bibfnamefont {S.}~\bibnamefont {Kucharski}}, \ and\ \bibinfo {author} {\bibfnamefont {J.}~\bibnamefont {Noga}},\ }\href {\doibase https://doi.org/10.1016/0009-2614(90)87031-L} {\bibfield  {journal} {\bibinfo  {journal} {Chemical Physics Letters}\ }\textbf {\bibinfo {volume} {165}},\ \bibinfo {pages} {513} (\bibinfo {year} {1990})}\BibitemShut {NoStop}%
\bibitem [{\citenamefont {Raghavachari}\ \emph {et~al.}(1989)\citenamefont {Raghavachari}, \citenamefont {Trucks}, \citenamefont {Pople},\ and\ \citenamefont {Head-Gordon}}]{Ref143}%
  \BibitemOpen
  \bibfield  {author} {\bibinfo {author} {\bibfnamefont {K.}~\bibnamefont {Raghavachari}}, \bibinfo {author} {\bibfnamefont {G.~W.}\ \bibnamefont {Trucks}}, \bibinfo {author} {\bibfnamefont {J.~A.}\ \bibnamefont {Pople}}, \ and\ \bibinfo {author} {\bibfnamefont {M.}~\bibnamefont {Head-Gordon}},\ }\href {\doibase https://doi.org/10.1016/S0009-2614(89)87395-6} {\bibfield  {journal} {\bibinfo  {journal} {Chemical Physics Letters}\ }\textbf {\bibinfo {volume} {157}},\ \bibinfo {pages} {479} (\bibinfo {year} {1989})}\BibitemShut {NoStop}%
\bibitem [{\citenamefont {He}\ \emph {et~al.}(2001)\citenamefont {He}, \citenamefont {He},\ and\ \citenamefont {Cremer}}]{Ref155}%
  \BibitemOpen
  \bibfield  {author} {\bibinfo {author} {\bibfnamefont {Y.}~\bibnamefont {He}}, \bibinfo {author} {\bibfnamefont {Z.}~\bibnamefont {He}}, \ and\ \bibinfo {author} {\bibfnamefont {D.}~\bibnamefont {Cremer}},\ }\href {\doibase 10.1007/s002140000196} {\bibfield  {journal} {\bibinfo  {journal} {Theoretical Chemistry Accounts: Theory, Computation, and Modeling (Theoretica Chimica Acta)}\ }\textbf {\bibinfo {volume} {105}},\ \bibinfo {pages} {182} (\bibinfo {year} {2001})}\BibitemShut {NoStop}%
\bibitem [{\citenamefont {Bishop}\ and\ \citenamefont {L\"uhrmann}(1978)}]{Ref69}%
  \BibitemOpen
  \bibfield  {author} {\bibinfo {author} {\bibfnamefont {R.~F.}\ \bibnamefont {Bishop}}\ and\ \bibinfo {author} {\bibfnamefont {K.~H.}\ \bibnamefont {L\"uhrmann}},\ }\href {\doibase 10.1103/PhysRevB.17.3757} {\bibfield  {journal} {\bibinfo  {journal} {Phys. Rev. B}\ }\textbf {\bibinfo {volume} {17}},\ \bibinfo {pages} {3757} (\bibinfo {year} {1978})}\BibitemShut {NoStop}%
\bibitem [{\citenamefont {March}\ \emph {et~al.}(1995)\citenamefont {March}, \citenamefont {Young},\ and\ \citenamefont {Sampanthar}}]{Ref13}%
  \BibitemOpen
  \bibfield  {author} {\bibinfo {author} {\bibfnamefont {N.}~\bibnamefont {March}}, \bibinfo {author} {\bibfnamefont {W.}~\bibnamefont {Young}}, \ and\ \bibinfo {author} {\bibfnamefont {S.}~\bibnamefont {Sampanthar}},\ }\href@noop {} {\emph {\bibinfo {title} {The Many-Body Problem in Quantum Physics}}}\ (\bibinfo  {publisher} {Dover Publications},\ \bibinfo {year} {1995})\BibitemShut {NoStop}%
\bibitem [{\citenamefont {Carr}\ and\ \citenamefont {Maradudin}(1964)}]{Ref91}%
  \BibitemOpen
  \bibfield  {author} {\bibinfo {author} {\bibfnamefont {W.~J.}\ \bibnamefont {Carr}}\ and\ \bibinfo {author} {\bibfnamefont {A.~A.}\ \bibnamefont {Maradudin}},\ }\href {\doibase 10.1103/PhysRev.133.A371} {\bibfield  {journal} {\bibinfo  {journal} {Phys. Rev.}\ }\textbf {\bibinfo {volume} {133}},\ \bibinfo {pages} {A371} (\bibinfo {year} {1964})}\BibitemShut {NoStop}%
\bibitem [{\citenamefont {Sawada}(1957)}]{Ref92}%
  \BibitemOpen
  \bibfield  {author} {\bibinfo {author} {\bibfnamefont {K.}~\bibnamefont {Sawada}},\ }\href {\doibase 10.1103/PhysRev.106.372} {\bibfield  {journal} {\bibinfo  {journal} {Phys. Rev.}\ }\textbf {\bibinfo {volume} {106}},\ \bibinfo {pages} {372} (\bibinfo {year} {1957})}\BibitemShut {NoStop}%
\bibitem [{\citenamefont {Lietz}(2019)}]{Ref5}%
  \BibitemOpen
  \bibfield  {author} {\bibinfo {author} {\bibfnamefont {J.~G.}\ \bibnamefont {Lietz}},\ }\href@noop {} {\enquote {\bibinfo {title} {Computational developments for ab initio many-body theory},}\ } (\bibinfo {year} {2019})\BibitemShut {NoStop}%
\bibitem [{\citenamefont {MacKay}(1992)}]{Ref161}%
  \BibitemOpen
  \bibfield  {author} {\bibinfo {author} {\bibfnamefont {D.~J.~C.}\ \bibnamefont {MacKay}},\ }\href {\doibase 10.1162/neco.1992.4.3.415} {\bibfield  {journal} {\bibinfo  {journal} {Neural Computation}\ }\textbf {\bibinfo {volume} {4}},\ \bibinfo {pages} {415} (\bibinfo {year} {1992})},\ \Eprint {http://arxiv.org/abs/https://direct.mit.edu/neco/article-pdf/4/3/415/812340/neco.1992.4.3.415.pdf} {https://direct.mit.edu/neco/article-pdf/4/3/415/812340/neco.1992.4.3.415.pdf} \BibitemShut {NoStop}%
\bibitem [{\citenamefont {Tipping}(2001)}]{Ref162}%
  \BibitemOpen
  \bibfield  {author} {\bibinfo {author} {\bibfnamefont {M.~E.}\ \bibnamefont {Tipping}},\ }\href {\doibase 10.1162/15324430152748236} {\bibfield  {journal} {\bibinfo  {journal} {J. Mach. Learn. Res.}\ }\textbf {\bibinfo {volume} {1}},\ \bibinfo {pages} {211–244} (\bibinfo {year} {2001})}\BibitemShut {NoStop}%
\bibitem [{\citenamefont {Zhang}\ and\ \citenamefont {Ling}(2018)}]{Ref17}%
  \BibitemOpen
  \bibfield  {author} {\bibinfo {author} {\bibfnamefont {Y.}~\bibnamefont {Zhang}}\ and\ \bibinfo {author} {\bibfnamefont {C.}~\bibnamefont {Ling}},\ }\href {\doibase 10.1038/s41524-018-0081-z} {\bibfield  {journal} {\bibinfo  {journal} {npj Computational Materials}\ }\textbf {\bibinfo {volume} {4}} (\bibinfo {year} {2018}),\ 10.1038/s41524-018-0081-z}\BibitemShut {NoStop}%
\bibitem [{\citenamefont {Baardsen}\ \emph {et~al.}(2013)\citenamefont {Baardsen}, \citenamefont {Ekstr\"om}, \citenamefont {Hagen},\ and\ \citenamefont {Hjorth-Jensen}}]{Ref8}%
  \BibitemOpen
  \bibfield  {author} {\bibinfo {author} {\bibfnamefont {G.}~\bibnamefont {Baardsen}}, \bibinfo {author} {\bibfnamefont {A.}~\bibnamefont {Ekstr\"om}}, \bibinfo {author} {\bibfnamefont {G.}~\bibnamefont {Hagen}}, \ and\ \bibinfo {author} {\bibfnamefont {M.}~\bibnamefont {Hjorth-Jensen}},\ }\href {\doibase 10.1103/PhysRevC.88.054312} {\bibfield  {journal} {\bibinfo  {journal} {Phys. Rev. C}\ }\textbf {\bibinfo {volume} {88}},\ \bibinfo {pages} {054312} (\bibinfo {year} {2013})}\BibitemShut {NoStop}%
\bibitem [{\citenamefont {Hagen}\ \emph {et~al.}(2014{\natexlab{a}})\citenamefont {Hagen}, \citenamefont {Papenbrock}, \citenamefont {Ekstr\"om}, \citenamefont {Wendt}, \citenamefont {Baardsen}, \citenamefont {Gandolfi}, \citenamefont {Hjorth-Jensen},\ and\ \citenamefont {Horowitz}}]{Ref9}%
  \BibitemOpen
  \bibfield  {author} {\bibinfo {author} {\bibfnamefont {G.}~\bibnamefont {Hagen}}, \bibinfo {author} {\bibfnamefont {T.}~\bibnamefont {Papenbrock}}, \bibinfo {author} {\bibfnamefont {A.}~\bibnamefont {Ekstr\"om}}, \bibinfo {author} {\bibfnamefont {K.~A.}\ \bibnamefont {Wendt}}, \bibinfo {author} {\bibfnamefont {G.}~\bibnamefont {Baardsen}}, \bibinfo {author} {\bibfnamefont {S.}~\bibnamefont {Gandolfi}}, \bibinfo {author} {\bibfnamefont {M.}~\bibnamefont {Hjorth-Jensen}}, \ and\ \bibinfo {author} {\bibfnamefont {C.~J.}\ \bibnamefont {Horowitz}},\ }\href {\doibase 10.1103/PhysRevC.89.014319} {\bibfield  {journal} {\bibinfo  {journal} {Phys. Rev. C}\ }\textbf {\bibinfo {volume} {89}},\ \bibinfo {pages} {014319} (\bibinfo {year} {2014}{\natexlab{a}})}\BibitemShut {NoStop}%
\bibitem [{\citenamefont {Baardsen}(2014)}]{Ref4}%
  \BibitemOpen
  \bibfield  {author} {\bibinfo {author} {\bibfnamefont {G.}~\bibnamefont {Baardsen}},\ }\href@noop {} {\enquote {\bibinfo {title} {Coupled-cluster theory for infinite matter},}\ } (\bibinfo {year} {2014})\BibitemShut {NoStop}%
\bibitem [{\citenamefont {Hagen}\ \emph {et~al.}(2014{\natexlab{b}})\citenamefont {Hagen}, \citenamefont {Papenbrock}, \citenamefont {Hjorth-Jensen},\ and\ \citenamefont {Dean}}]{Ref16}%
  \BibitemOpen
  \bibfield  {author} {\bibinfo {author} {\bibfnamefont {G.}~\bibnamefont {Hagen}}, \bibinfo {author} {\bibfnamefont {T.}~\bibnamefont {Papenbrock}}, \bibinfo {author} {\bibfnamefont {M.}~\bibnamefont {Hjorth-Jensen}}, \ and\ \bibinfo {author} {\bibfnamefont {D.~J.}\ \bibnamefont {Dean}},\ }\href {\doibase 10.1088/0034-4885/77/9/096302} {\bibfield  {journal} {\bibinfo  {journal} {Reports on Progress in Physics}\ }\textbf {\bibinfo {volume} {77}},\ \bibinfo {pages} {096302} (\bibinfo {year} {2014}{\natexlab{b}})}\BibitemShut {NoStop}%
\bibitem [{\citenamefont {Hu}\ \emph {et~al.}(2022)\citenamefont {Hu}, \citenamefont {Jiang}, \citenamefont {Miyagi}, \citenamefont {Sun}, \citenamefont {Ekström}, \citenamefont {Forssén}, \citenamefont {Hagen}, \citenamefont {Holt}, \citenamefont {Papenbrock}, \citenamefont {Stroberg},\ and\ \citenamefont {Vernon}}]{Ref20}%
  \BibitemOpen
  \bibfield  {author} {\bibinfo {author} {\bibfnamefont {B.}~\bibnamefont {Hu}}, \bibinfo {author} {\bibfnamefont {W.}~\bibnamefont {Jiang}}, \bibinfo {author} {\bibfnamefont {T.}~\bibnamefont {Miyagi}}, \bibinfo {author} {\bibfnamefont {Z.}~\bibnamefont {Sun}}, \bibinfo {author} {\bibfnamefont {A.}~\bibnamefont {Ekström}}, \bibinfo {author} {\bibfnamefont {C.}~\bibnamefont {Forssén}}, \bibinfo {author} {\bibfnamefont {G.}~\bibnamefont {Hagen}}, \bibinfo {author} {\bibfnamefont {J.~D.}\ \bibnamefont {Holt}}, \bibinfo {author} {\bibfnamefont {T.}~\bibnamefont {Papenbrock}}, \bibinfo {author} {\bibfnamefont {S.~R.}\ \bibnamefont {Stroberg}}, \ and\ \bibinfo {author} {\bibfnamefont {I.}~\bibnamefont {Vernon}},\ }\href {\doibase 10.1038/s41567-022-01715-8} {\bibfield  {journal} {\bibinfo  {journal} {Nature Physics}\ }\textbf {\bibinfo {volume} {18}} (\bibinfo {year} {2022}),\ 10.1038/s41567-022-01715-8}\BibitemShut {NoStop}%
\bibitem [{\citenamefont {Hagen}\ \emph {et~al.}(2016)\citenamefont {Hagen}, \citenamefont {Hjorth-Jensen}, \citenamefont {Jansen},\ and\ \citenamefont {Papenbrock}}]{Ref43}%
  \BibitemOpen
  \bibfield  {author} {\bibinfo {author} {\bibfnamefont {G.}~\bibnamefont {Hagen}}, \bibinfo {author} {\bibfnamefont {M.}~\bibnamefont {Hjorth-Jensen}}, \bibinfo {author} {\bibfnamefont {G.~R.}\ \bibnamefont {Jansen}}, \ and\ \bibinfo {author} {\bibfnamefont {T.}~\bibnamefont {Papenbrock}},\ }\href {\doibase 10.1088/0031-8949/91/6/063006} {\bibfield  {journal} {\bibinfo  {journal} {Physica Scripta}\ }\textbf {\bibinfo {volume} {91}},\ \bibinfo {pages} {063006} (\bibinfo {year} {2016})}\BibitemShut {NoStop}%
\bibitem [{\citenamefont {Hergert}\ \emph {et~al.}(2016)\citenamefont {Hergert}, \citenamefont {Bogner}, \citenamefont {Morris}, \citenamefont {Schwenk},\ and\ \citenamefont {Tsukiyama}}]{Ref48}%
  \BibitemOpen
  \bibfield  {author} {\bibinfo {author} {\bibfnamefont {H.}~\bibnamefont {Hergert}}, \bibinfo {author} {\bibfnamefont {S.}~\bibnamefont {Bogner}}, \bibinfo {author} {\bibfnamefont {T.}~\bibnamefont {Morris}}, \bibinfo {author} {\bibfnamefont {A.}~\bibnamefont {Schwenk}}, \ and\ \bibinfo {author} {\bibfnamefont {K.}~\bibnamefont {Tsukiyama}},\ }\href {\doibase https://doi.org/10.1016/j.physrep.2015.12.007} {\bibfield  {journal} {\bibinfo  {journal} {Physics Reports}\ }\textbf {\bibinfo {volume} {621}},\ \bibinfo {pages} {165} (\bibinfo {year} {2016})},\ \bibinfo {note} {memorial Volume in Honor of Gerald E. Brown}\BibitemShut {NoStop}%
\bibitem [{\citenamefont {Morris}\ \emph {et~al.}(2015)\citenamefont {Morris}, \citenamefont {Parzuchowski},\ and\ \citenamefont {Bogner}}]{Ref65}%
  \BibitemOpen
  \bibfield  {author} {\bibinfo {author} {\bibfnamefont {T.~D.}\ \bibnamefont {Morris}}, \bibinfo {author} {\bibfnamefont {N.~M.}\ \bibnamefont {Parzuchowski}}, \ and\ \bibinfo {author} {\bibfnamefont {S.~K.}\ \bibnamefont {Bogner}},\ }\href {\doibase 10.1103/PhysRevC.92.034331} {\bibfield  {journal} {\bibinfo  {journal} {Phys. Rev. C}\ }\textbf {\bibinfo {volume} {92}},\ \bibinfo {pages} {034331} (\bibinfo {year} {2015})}\BibitemShut {NoStop}%
\end{thebibliography}%

\end{document}